\documentclass[aps,preprint]{revtex4}

\usepackage{graphics,graphicx}
\usepackage{amsmath,amssymb}
\usepackage{epsfig}
\usepackage{txfonts}
\usepackage{subfigure}
\usepackage{xcolor}
\usepackage{epstopdf}
\usepackage{hyperref}
\begin{document}

\title{Flat and bent branes with inner structure in two-field mimetic gravity}
\author{Qian Xiang $^{1}$\footnote{Email:xiangq18@lzu.edu.cn}}
\author{Yi Zhong $^{2}$\footnote{Email:zhongy@hnu.edu.cn}}
\author{Qun-Ying Xie $^{1,3}$\footnote{Email:xieqy@lzu.edu.cn} }
\author{Li Zhao $^{1}$\footnote{Email:lizhao@lzu.edu.cn, corresponding author}}
\affiliation{\small{$^{1}$ Institute of Theoretical Physics, Lanzhou University,  Lanzhou 730000, China\\
                    $^{2}$ School of Physics and Electronics Science, Hunan University, Changsha 410082,  China\\
                  $^{3}$ School of Information Science and Engineering, Lanzhou University, Lanzhou 730000, China\\\\}}

\begin{abstract}
Inspired by the  work [Eur. Phys. J. C 78 (2018) 45], we study the linear  tensor perturbation  of  both the flat and bent thick branes with inner structure in two-field mimetic gravity. The master equations for the linear tensor perturbations are derived by taking the transverse and traceless gauge.  For the Minsowskii and Anti-de-Sitter brane, the brane systems are stable against the tensor perturbation.  The effective potentials
of the tensor perturbations of both the flat and bent thick branes are  volcano-like, and this structure may potentially lead to the zero-mode  and the resonant  modes  of the tensor perturbation.  We further illustrate the results of massive resonant modes.
\end{abstract}


\maketitle

\section{Introduction}

The braneworld theory has received much attention
over the past years since it proposes a new route to possibly solve the gauge hierarchy problem and the cosmological constant problem  \cite{Arkani-Hamed1998,Antoniadis1998,RandallS1999,Randall1999}. The Randall-Sundrum (RS)
models \cite{RandallS1999,Randall1999} are the typical examples of them with  a non-factorizable  metric and a warped extra dimension. It is shown that RS models give rise to thin brane profiles because the warp factor  has cusp singularities at the brane positions. Several proposals for generalizing thin branes into  thick branes have been presented in the literature. The thick branes  are obtained by introducing one or  more bulk scalar fields  coupled to gravity \cite{Gremm:1999pj,DeWolfe:1999cp,Kobayashi:2001jd,Bazeia:2002sd,Wang:2002pka,Bazeia:2004dh,Afonso:2006gi,Bazeia:2006ef,Bogdanos:2006qw,
Dzhunushaliev:2009va,Liu:2009ega}, or  realized by pure geometric frameworks without background matter fields considered in \cite{Rubakov:1983rsd,BarbosaCendejas:2005kn,HerreraAguilar:2010kt,Zhong:2015pta}. Most of the models study  Minkowski branes, and  a few  of them consider  the curvature of the embedded brane, which includes de Sitter ($\text{dS}$) or
anti-de Sitter ($\text{AdS}$) geometry.

On the other hand, mimetic gravity is proposed by Chamseddine and Mukhanov \cite{Chamseddine2013}  as one of the extensions of  general relativity (GR). In the original setup, a physical metric $g_{\mu\nu}$  is defined in terms of an auxiliary metric  $\tilde{g}_{\mu\nu}$ and a scalar field $\phi$ with the  relation $g_{\mu\nu}=-\tilde{g}_{\mu\nu}g^{\alpha\beta}\partial_{\alpha}\phi\partial_{\beta}\phi$. This model regards the scalar field as the conformal degree of freedom  to mimick cold dark matter.  The mimetic model is then investigated by adding a  potential $V(\phi)$ of the scalar field to explain the cosmological issues \cite{Chamseddine2014,Lim2010}.   With the appropriate choice of the potential, it is possible to  provide an inflationary mechanism and a bouncing universe within this framework \cite{Chamseddine2014}. Therefore, this stimulates some interests in phenomenology and its
observational viability \cite{Babichev2017,Sadeghnezhad2017,Casalino2018,Ganz2019,Zheng2017,Myrzakulov:2015kda,Vagnozzi:2017ilo,Casalino:2018wnc}, and leads to the  Hamiltonian analyses of different mimetic models \cite{Malaeb2015,Chaichian2014,Takahashi2017,Zheng:2018cuc,Shen:2019nyp}.   Mimetic gravity in various modified gravity theories
has also been widely discussed in \cite{Nojiri:2014zqa,Nojiri:2016vhu,Nojiri:2016ppu,Odintsov:2015wwp,Odintsov:2015cwa,Astashenok:2015haa,Nojiri:2017ygt,Cognola2016,Hosseinkhan2018,Guo2020,Chen:2020zzs}.
Another interest in mimetic gravity is to investigate the mimetic gravity in braneworld scenarios. The
late-time cosmic expansion and inflation  are investigated in the mimetic RS braneworld \cite{Sadeghnezhad2017}. Following this work,  the late time acceleration and perturbation behavior are studied for the brane-anti-brane system \cite{Sadatian2019}. Later, the tensor and
scalar perturbations on several thick branes are investigated in mimetic gravity   \cite{Yi2018}.

As we know,  the scalar fields are usually introduced to generate the topological defects  such as kinks and domain walls for realizing the thick branes,  so it is natural to  generate the  domain walls by the mimetic scalar fields. However, suffering from the ghost and gradient instabilities, the original mimetic theory with single field could not suffice \cite{Firouzjahi:2017txv}. For the single field mimetic scenario, one can go to a ghost-free theory by adding higher derivative terms to the original theory \cite{Gorji:2017cai}. For the two-field extension of the mimetic gravity  put forward in  \cite{Firouzjahi2018}, the double scalar fields version not only avoids the above problem but also allows us to construct the thick branes with a complicated inner structure.
It is common to find the background solution of the thick branes via the first-order formalism \cite{Afonso:2006gi} or the extension method  \cite{Bazeia:2013uba}.
 However, in the two-field mimetic theory,  the gravity  and two mimetic scalars are  coupled. It is relatively harder to find the domain wall solutions  if the four-dimensional  geometry is either flat or curved.
 In this work, we apply the reconstruction
technique \cite{Higuchi:2014hn} to seek the background solutions of  three cases of thick branes. The approach gives the form of the warp factor and scalar fields, and provides a direct way to investigate  the other variables. We consider that the thick  domain walls  possess four-dimensional $\text{dS}$ and $\text{AdS}$ symmetries  as well as the Poincar\'{e} one.   Because $\text{dS}$ and $\text{AdS}$ branes have different inner structures from  flat  brane, it is of interest to consider the tensor perturbation of gravity on the bent branes. 

 The paper is organized as follows. In section \ref{sec:model}, we introduce the  mimetic
thick brane model and consider three cases of thick branes. In section \ref{Sec3}, we analyze the stability of the model under the linear tensor perturbation and study the localization of gravity zero-mode. Finally, a brief conclusion is presented in Section \ref{Sec4}. Throughout this paper, capital Latin letters $M,\;N$,... represent the five-dimensional coordinate indices running over $0,1,2,3,5$, and  lower-case Greek letters $\mu,\;\nu$,... represent the four-dimensional coordinate indices running over $0,1,2,3$.

\section{Brane Setup and field equations}
\label{sec:model}
We consider the  five-dimensional two-field mimetic gravity  where the action  is the Einstein-Hilbert action
constructed in terms of the physical metric $g_{MN}$.
For our model, in the natural unit, the action can be written as a  Lagrange multiplier formulation,
\begin{eqnarray}
        S\!=\!\int d^4x dy\sqrt{-g}\left( \frac{R}{2}+ \mathcal{L}_{m} \right),
        \label{action mgb1}
    \end{eqnarray}
 where the Lagrangian of the two interacting  mimetic scalar field is generalized as,
 \begin{eqnarray}
 \mathcal{L}_{m}=\lambda\left[g^{MN}\partial_M \phi_{1} \partial_N \phi_{1}+g^{MN}\partial_M \phi_{2} \partial_N \phi_{2}-U(\phi_{1},\phi_{2})\right]-V(\phi_{1},\phi_{2}).
 \end{eqnarray}
 In  the original  mimetic model $U(\phi)=-1$, and here the potential is extended into  the form  of $U(\phi_{1},\phi_{2})$ with double mimetic fields. The Lagrange multiplier $\lambda$ enforces the mimetic constraint,
\begin{equation}
g^{MN}\partial_M \phi_{1} \partial_N \phi_{1}+g^{MN}\partial_M \phi_{2} \partial_N \phi_{2}-U(\phi_{1},\phi_{2})=0.\label{constraint1}
\end{equation}
The variation of the action (\ref{action mgb1}) concerning the metric $g_{MN}$ and the two scalar fields ($\phi_{1}$ and $\phi_{2}$) yields the following field equations, respectively:
\begin{eqnarray}
        \label{var eom1}
        G_{MN}+2\lambda \partial_M \phi_{1} \partial_N \phi_{1}+2\lambda \partial_M \phi_{2} \partial_N \phi_{2}-\mathcal{L}_{m}g_{MN}=0, \nonumber   \\
        \label{var eom2}
        2\lambda\Box^{(5)}\phi_{1}+2\nabla_{M}\lambda\nabla^{M}\phi_{1}+\lambda \frac{\partial U(\phi_{1},\phi_{2})}{\partial \phi_{1}}+\frac{\partial V(\phi_{1},\phi_{2})}{\partial \phi_{1}}=0, \nonumber\\
        \label{var eom3}
        2\lambda\Box^{(5)}\phi_{2}+2\nabla_{M}\lambda\nabla^{M}\phi_{2}+\lambda \frac{\partial U(\phi_{1},\phi_{2})}{\partial \phi_{2}}+\frac{\partial V(\phi_{1},\phi_{2})}{\partial \phi_{2}}=0.\label{EoM}
    \end{eqnarray}
The line-element for a warped five-dimensional geometry is generally assumed as,
\begin{eqnarray}
        \label{brane metric1}
       ds^2=a^2(y)\hat{g}_{\mu\nu}dx^{\mu}dx^{\nu}+dy^2
    \end{eqnarray}
with $y$ the extra spatial coordinate. We deal with $a=a(y)$,  $\phi_{1}=\phi_{1}(y)$ and $\phi_{2}=\phi_{2}(y)$ when considering the static brane.
    The metrics $\hat{g}_{\mu\nu}$ on the branes reads,
\begin{eqnarray}
    \hat{g}_{\mu\nu}=\left\{
    \begin{array}{cc}
    -dt^2+(dx_1^2+dx_2^2+dx_3^2)~~~\textrm{$M_{4}$ brane},\label{Minkowski}\\
    -dt^2+e^{2\sqrt{\Lambda_4} t}(dx_1^2+dx_2^2+dx_3^2)~~~\textrm{$\text{dS}_4$ brane},\label{ds}\\
    e^{-2\sqrt{-\Lambda_4}x_3}(-dt^2+dx_1^2+dx_2^2)+dx_3^2~~~\textrm{$\text{AdS}_4$ brane},\label{Ads}\\
    \end{array}\right.~~
\end{eqnarray}
where $\Lambda_4$ is related to the  four-dimensional cosmological constant of  dS$_4$ or AdS$_4$ brane \cite{Liu:2009dt,Liu:2011zy}.

\begin{figure}[!htb]
    \begin{center}
    \subfigure[The warp factor]{\label{fig model 1}
        \includegraphics[width=2.0in]{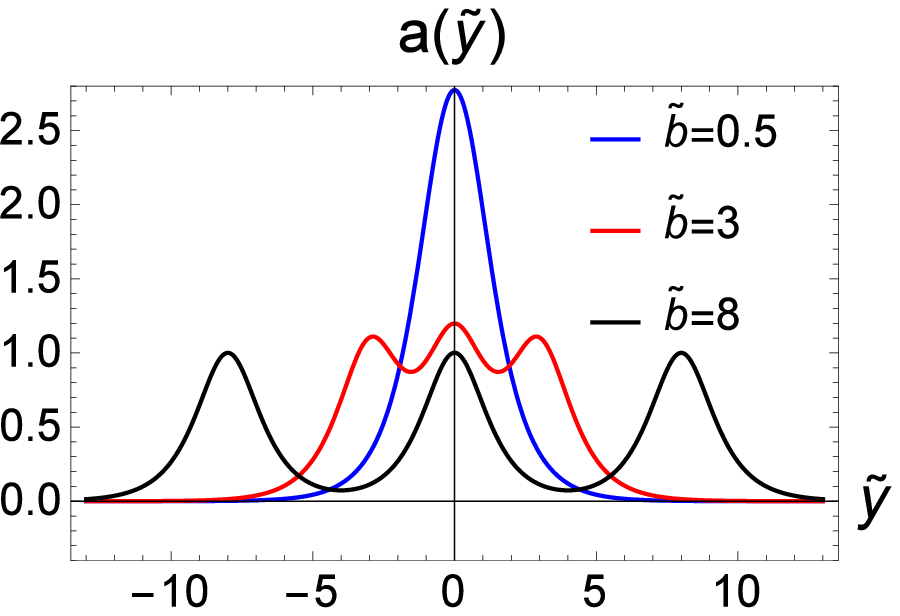}}
     \subfigure[The scalar field $\phi_{1}(\tilde{y})$]{\label{fig model 2}
     \includegraphics[width=2.0in]{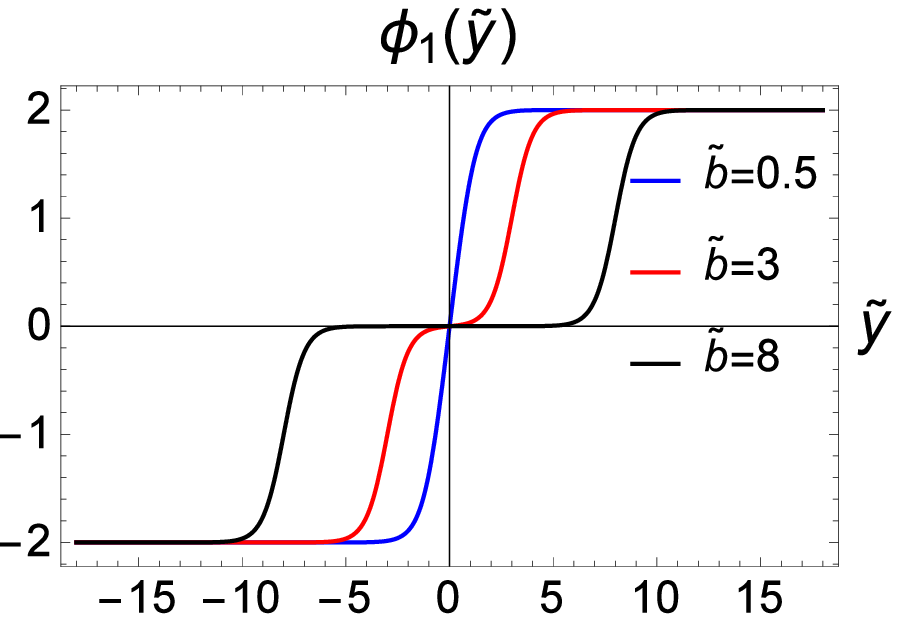}}
     \subfigure[The scalar field $\phi_{2}(\tilde{y})$]{\label{fig model 3}
     \includegraphics[width=2.0in]{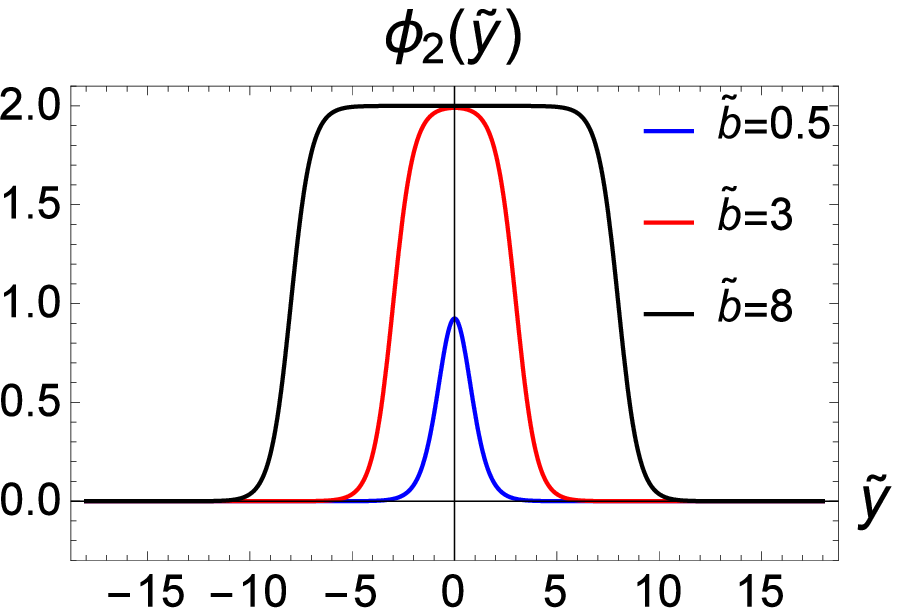}}
    \end{center}
    \caption{Plots of the warp factor $a(\tilde{y})$ and two mimetic scalar fields $\phi_{1}(\tilde{y})$ and $\phi_{2}(\tilde{y})$ of the brane models. }
    \label{warpfactorandscalarfield}
    \end{figure}

We know that Eqs.(\ref{constraint1}) and (\ref{EoM}) determine the solution of the brane system. There are three independent equations and  six variables in this system. To solve this system, we should preset three variables.
Based on the reconstruction technique, we will give the form of $a(y)$, $\phi_{1}(y)$, $\phi_{2}(y)$ and try to find the solution of $U(\phi_{1},\phi_{2})$, $V(\phi_{1},\phi_{2})$ and $\lambda(\phi_{1},\phi_{2})$ . Here  the warp factor $a(y)$ and the scalar fields $\phi_{1}(y)$, $\phi_{2}(y)$  are  given by
\begin{eqnarray}
        a(y)\!\!&=\!\!& \text{sech}(k (y-b))+\text{sech}(k y)+\text{sech}(k (y+b)), \label{warpfactor} \\
              \phi_{1}(y)\!\!&=\!\!& \text{tanh}(k(y-b))+\text{tanh}(k(y+b)),\label{scalarfield1} \\
              \phi_{2}(y)\!\!&=\!\!& \text{tanh}(k(y-b))-\text{tanh}(k(y+b))\label{scalarfield2},
    \end{eqnarray}
where $k$ and $b$ are  parameters with dimension mass and  length, respectively. To display the
configuration of $ a(y)$, $\phi_{1}(y)$ and $\phi_{2}(y)$, we introduce the dimensionless quantities $\tilde{y}=ky$ and $\tilde{b}=kb$.  The shapes of the warp factor and the two scalar fields are depicted in Fig. \ref{warpfactorandscalarfield}. The above warp factor indicates that the bulk space-time is asymptotically $\text{AdS}$, which is essential for the localization of gravitation. The brane  splits from a single brane into three sub-branes as the parameter $\tilde{b}$ increases. Thus, these branes have a rich inner structure. The scalar field $\phi_{1}(\tilde{y})$ supports the topological  solution  which changes from a single-kink to a double-kink configuration with the increasing of $\tilde{b}$.  Without loss of generality, the other scalar field $\phi_{2}(\tilde{y})$  is assumed as a non-topologically lump-like solution.

\subsection{Flat brane}
For Minkowski brane, we take $\Lambda_{4}=0$. With the ansatz (\ref{Minkowski}), the Einstein tensors  are expressed as
\begin{eqnarray}
G_{\mu\nu}=3\eta_{\mu\nu} \left(a(y) a''(y)+a'(y)^2\right),\;\; G_{55}=6a'(y)^2/a(y)^{2}
 \end{eqnarray}
Eqs. (\ref{EoM}) and (\ref{constraint1})  can be reduced to
  \begin{eqnarray}
        \label{eom21}
      \frac{3a'^2}{a^2}+\frac{3a''}{a}+V(\phi_{1},\phi_{2})+\lambda \left(U(\phi_{1},\phi_{2})-\phi_{1}'^2-\phi_{2}'^2\right)=0, \\
       \label{eom22}
       \frac{6a'^2}{a^2}+V(\phi_{1},\phi_{2})+\lambda \left(U(\phi_{1},\phi_{2})+\phi_{1}'^2+\phi_{1}'^2\right)=0, \\
       \label{eom23}
        \frac{8\lambda a'\phi_{1}'}{a}+2\lambda\phi_{1}''+2\lambda'\phi_{1}'+\lambda\frac{\partial U(\phi_{1},\phi_{2})}{\partial \phi_{1}}
        +\frac{\partial V(\phi_{1},\phi_{2})}{\partial \phi_{1}}=0,  \\
       \label{eom24}
       \frac{8\lambda a'\phi_{2}'}{a}+2\lambda\phi_{2}''+2\lambda'\phi_{2}'+\lambda\frac{\partial U(\phi_{1},\phi_{2})}{\partial \phi_{2}}
        +\frac{\partial V(\phi_{1},\phi_{2})}{\partial \phi_{2}}=0,  \\
      \label{eom25}
       \phi_{1}'^2+\phi_{2}'^2=U(\phi_{1},\phi_{2}),\label{EoMS}
    \end{eqnarray}
where the prime denotes the derivative with respect to $y$. Plugging Eqs. (\ref{warpfactor}), (\ref{scalarfield1}) and (\ref{scalarfield2}) into (\ref{EoMS}),  we get the following analytic  solution,
\begin{eqnarray}
U(\phi_{1}(y),\phi_{2}(y))&=&2 k^2 \left[\text{sech}^4(k (y-b))+\text{sech}^4(k (y+b))\right],\nonumber \\
V(\phi_{1}(y),\phi_{2}(y))&=&\frac{3 k^2}{\bigg[\text{sech}(k (y-b))+\text{sech}(k (y+b))+\text{sech}(k y)\bigg]^2} \bigg[(\text{sech}(k (y-b)) \nonumber\\ &&+\text{sech}(k (y+b))+\text{sech}(k y))  (\text{sech}^3(k (y-b))-\frac{1}{2} (\cosh (2 k (y+b))-3)\nonumber \\&& \times \text{sech}^3(k (y+b)) -\tanh ^2(k (y-b)) \text{sech}(k (y-b))+\text{sech}^3(k y)-\tanh ^2(k y) \nonumber \\ && \times\text{sech}(k y))-(\tanh (k (y-b)) \text{sech}(k (y-b))+\tanh (k (y+b)) \text{sech}(k (y+b))\nonumber \\ &&+\tanh (k y) \text{sech}(k y))^2\bigg],\nonumber \\
\lambda(\phi_{1}(y),\phi_{2}(y))&=&\frac{-3}{8\bigg[\text{sech}(k (y-b))+\text{sech}(k (y+b))+\text{sech}(k y)\bigg]^2 \bigg[\text{sech}^4(k (y-b))+\text{sech}^4(k (y+b))\bigg]}\nonumber \\ && \times \bigg[2 \text{sech}^4(k (y-b))+2 \text{sech}^4(k (y+b))+2 \text{sech}(k y) \text{sech}^3(k (y-b))\nonumber \\&&+(\cosh (2 k (y-b))+\cosh (2 k (y+b))-\cosh (4 b k)+3) \text{sech}^3(k (y-b))\nonumber \\&& \times\text{sech}^3(k (y+b))+\text{sech}^3(k y) (2 \text{sech}(k (y-b))+(\cosh (2 k (y+b))-\cosh (2 b k)\nonumber \\&&+\cosh (2 k y)+3) \text{sech}^3(k (y+b))-2 \sinh ^2(b k) \text{sech}^3(k (y-b)))+2 \text{sech}^4(k y)\bigg].
\end{eqnarray}

\subsection{Bent branes}
We now turn attention to the case of $\Lambda_{4}\neq0$.  The presence of  $\Lambda_{4}$ makes the field equation and background solution complex. For the $\text{dS}_{4}$ geometry ($\Lambda_{4}>0$), with the ansatz (\ref{ds}), the Einstein tensors  are expressed as
\begin{eqnarray}
G_{\mu\nu}&=&-\frac{3 (\Lambda_{4}-(a'(y))^{2}-a(y)a''(y))}{a^{2}(y)}a^{2}(y)\hat{g}_{\mu\nu}, \nonumber\\
G_{55}&=&\frac{-6(\Lambda_{4}-a'(y)^{2})}{a^{2}(y)}.
\end{eqnarray}
Eqs. (\ref{EoM}) and (\ref{constraint1})  can be reduced to
  \begin{eqnarray}
        \label{eom21}
      -\frac{3 (\Lambda_{4}-(a'(y))^{2}-a(y)a''(y))}{a^{2}(y)}+V(\phi_{1},\phi_{2})+\lambda \left(U(\phi_{1},\phi_{2})-\phi_{1}'^2-\phi_{2}'^2\right)=0, \\ \label{eom22}
       \frac{-6(\Lambda_{4}-a'(y)^{2})}{a^{2}(y)}+V(\phi_{1},\phi_{2})+\lambda \left(U(\phi_{1},\phi_{2})+\phi_{1}'^2+\phi_{2}'^2\right)=0, \\
       \label{eom23}
        \frac{8\lambda a'(y)\phi_{1}'}{a(y)}+2\lambda\phi_{1}''+2\lambda'\phi_{1}'+\lambda\frac{\partial U(\phi_{1},\phi_{2})}{\partial \phi_{1}}
        +\frac{\partial V(\phi_{1},\phi_{2})}{\partial \phi_{1}}=0,  \\
       \label{eom24}
       \frac{8\lambda a'(y)\phi_{2}'}{a(y)}+2\lambda\phi_{2}''+2\lambda'\phi_{2}'+\lambda\frac{\partial U(\phi_{1},\phi_{2})}{\partial \phi_{2}}
        +\frac{\partial V(\phi_{1},\phi_{2})}{\partial \phi_{2}}=0,  \\
      \label{eom25}
       \phi_{1}'^2+\phi_{2}'^2=U(\phi_{1},\phi_{2}).
    \end{eqnarray}
Now  the system can be solved as,
\begin{eqnarray}
U(\phi_{1}(y),\phi_{2}(y))&=& 2 k^2 \left[\text{sech}^4(k (y-b))+\text{sech}^4(k (y+b))\right],\nonumber\\ \label{eom22}
V(\phi_{1}(y),\phi_{2}(y))&=& \frac{3}{\bigg[\text{sech}(k (y-b))+\text{sech}(k (y+b))+\text{sech}(k y)\bigg]^2}\bigg[\Lambda_{4} -k^2 (\tanh (k (y-b))\nonumber\\&& \times\text{sech}(k (y-b))+\tanh (k (y+b)) \text{sech}(k (y+b))+\tanh (k y) \text{sech}(k y))^2\nonumber\\&&+k^2 (\text{sech}(k (y-b))+\text{sech}(k (y+b))+\text{sech}(k y))\nonumber\\&& \times(\text{sech}^3(k (y-b))-\frac{1}{2} (\cosh (2 k (y+b))-3) \text{sech}^3(k (y+b))\nonumber\\&&-\tanh ^2(k (y-b)) \text{sech}(k (y-b))+\text{sech}^3(k y)-\tanh ^2(k y) \text{sech}(k y))\bigg],\nonumber\\
\lambda(\phi_{1}(y),\phi_{2}(y))&=&\frac{-3 }{8 k^2 \bigg[\text{sech}(k (y-b))+\text{sech}(k (y+b))+\text{sech}(k y)\bigg]^2 \bigg[\text{sech}^4(k (y-b))+\text{sech}^4(k (y+b))\bigg]}\nonumber\\ &\times&\bigg[-2\Lambda_{4}+2 k^2 \text{sech}^4(k (y-b))+2 k^2 \text{sech}^4(k (y+b))+2 k^2 \text{sech}(k y) \text{sech}^3(k (y-b)) \nonumber\\&+&k^2 (\cosh (2 k (y-b))+\cosh (2 k (y+b))-\cosh (4 b k)+3) \text{sech}^3(k (y-b)) \text{sech}^3(k (y+b))\nonumber\\&+&k^2 \text{sech}^3(k y) (2 \text{sech}(k (y-b))+(\cosh (2 k (y+b))-\cosh (2 b k)+\cosh (2 k y)+3) \nonumber\\& \times&\text{sech}^3(k (y+b))-2 \sinh ^2(b k) \text{sech}^3(k (y-b)))+2 k^2 \text{sech}^4(k y)\bigg].
\end{eqnarray}
For AdS case ($\Lambda_{4}<0$), we change $\Lambda_{4}\rightarrow -\Lambda_{4}$,  the solution of $\text{dS}_{4}$ brane is transformed into that of   $\text{AdS}_{4}$ brane.  This result is interesting, since it simplifies the calculation significantly.

\begin{figure}[htb]
\begin{center}
\includegraphics[width=2.0in]{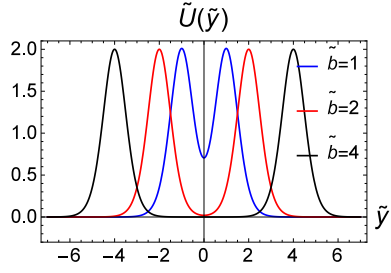}
\end{center}
\caption{The potential  $\tilde{U}(\tilde{y})$ for the three  cases of branes.} \label{subfig:phin}
\end{figure}

\begin{figure}[htb]
\begin{center}
\subfigure[$\text{M}_{4}$ brane]{
\includegraphics[width=2.0in]{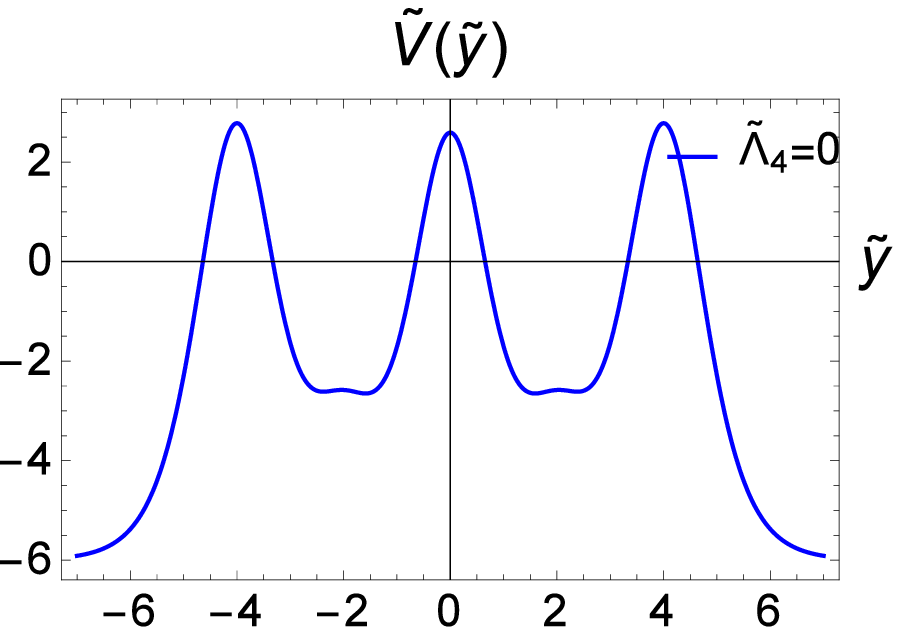}}
\subfigure[$\text{dS}_{4}$ brane]{
\includegraphics[width=2.0in]{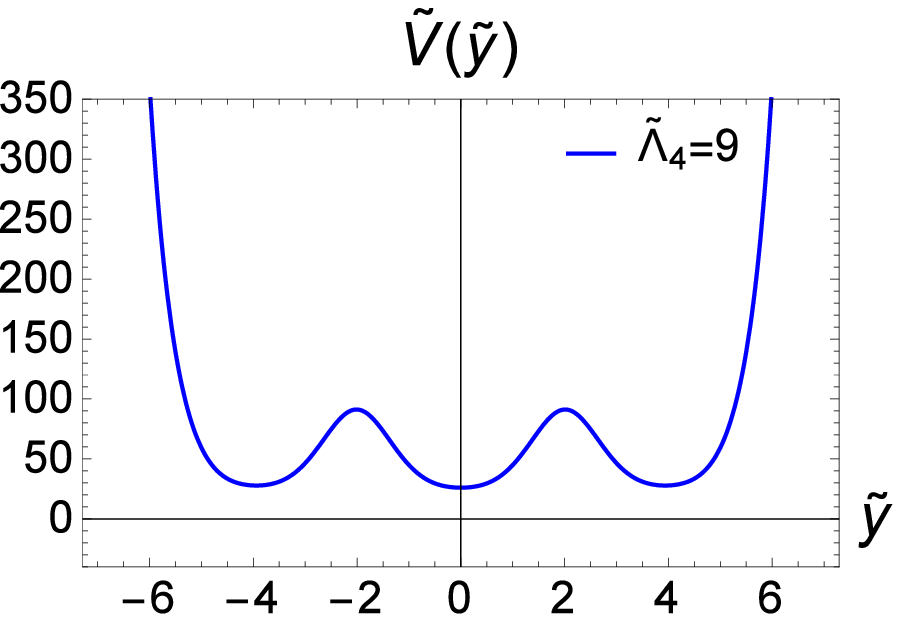}}
\subfigure[$\text{AdS}_{4}$ brane]{
\includegraphics[width=2.0in]{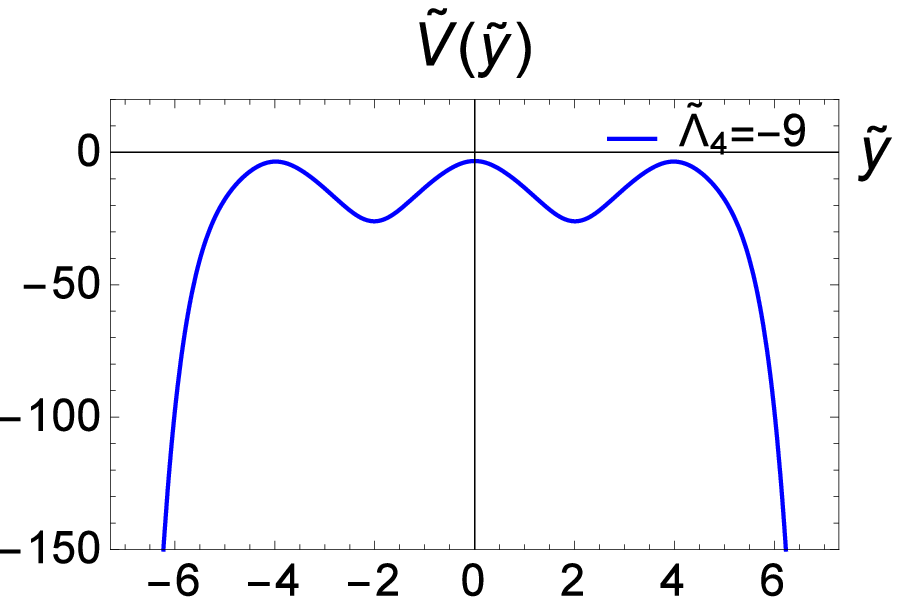}}
\end{center}
\caption{Comparison of the dimensionless scalar potential $\tilde{V}(\tilde{y})$ in terms of $\tilde{y}$. The parameter $\tilde{b}$ is set as $\tilde{b}=4$.}
\label{potential}
\end{figure}

\begin{figure}[!htb]
    \begin{center}
    \subfigure[$\text{M}_{4}$ brane]{\label{fig model 1}
        \includegraphics[width=2.0in]{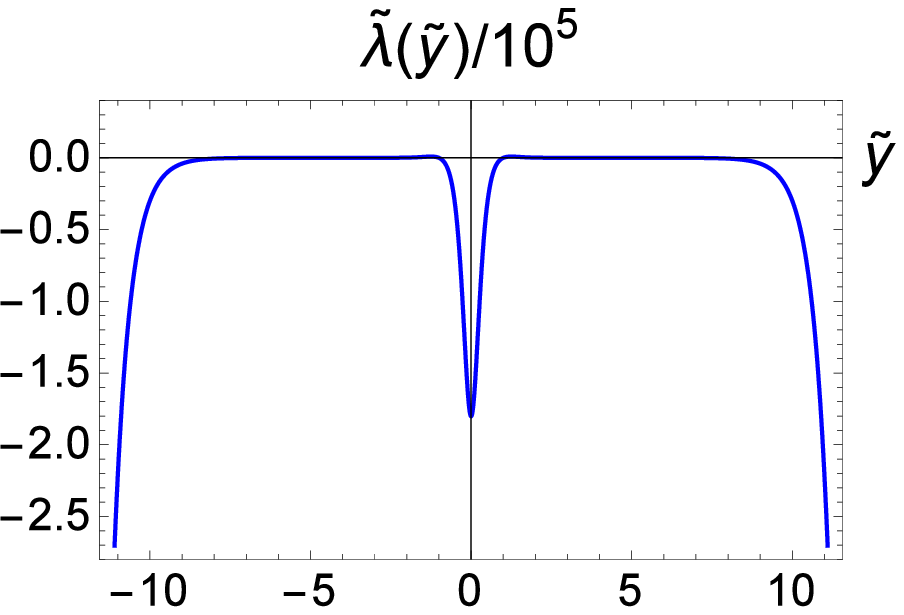}}
     \subfigure[$\text{dS}_{4}$ brane]{\label{fig model 2}
     \includegraphics[width=2.0in]{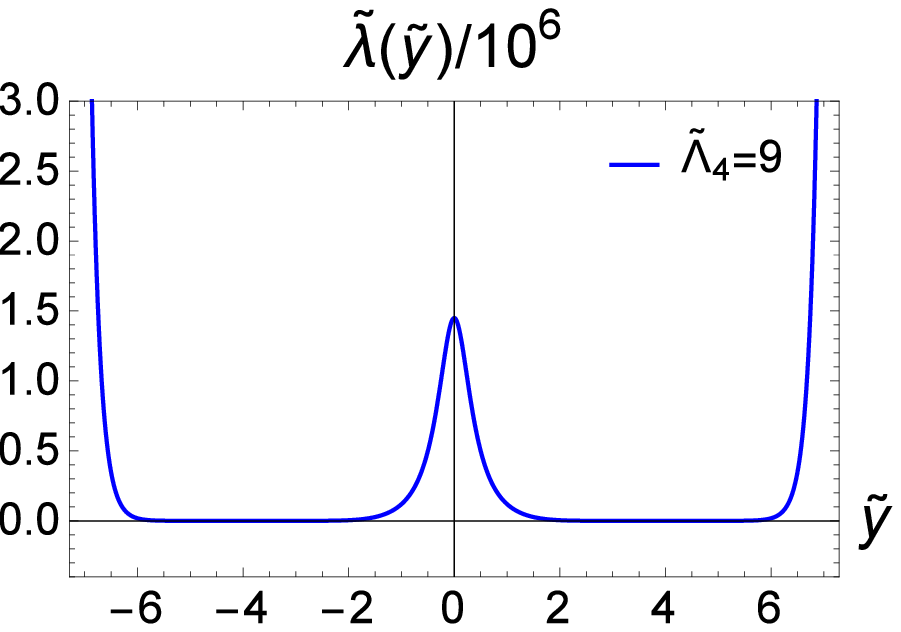}}
     \subfigure[$\text{AdS}_{4}$ brane]{\label{fig model 2}
     \includegraphics[width=2.0in]{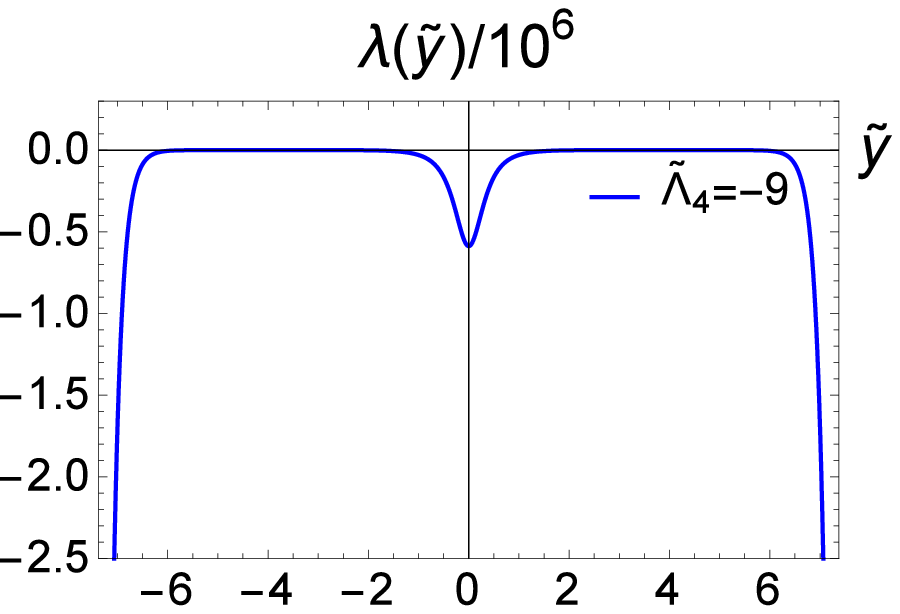}}
    \end{center}
    \caption{Comparison of Lagrange multiplier $\tilde{\lambda}(\tilde{y})$  with the parameter $\tilde{b}=4$.}
    \label{lamda}
    \end{figure}

So far, we have obtained the background solution of  the three cases of  thick branes. In more detail, the potential $U(y)$ is the same; however, $V(y)$ and $\lambda(y)$ take different values for they are related to the parameter  $\Lambda_{4}$.
 For the sake of clarity, by performing the rescaling of the quantities,
\begin{equation}
\tilde{y}=ky, \tilde{b}=kb, \tilde{\Lambda}_{4}=\Lambda_{4}/k^{2}, \tilde{V}(\tilde{y})=V(\tilde{y})/k^{2}, \tilde{U}(\tilde{y})=U(\tilde{y})/k^{2}, \tilde{\lambda}(\tilde{y})=\lambda(\tilde{y}),
 \end{equation}
 we can obtain the dimensionless quantities. The profiles of the potential  $\tilde{U}(\tilde{y})$  with the increasing of the parameters $\tilde{b}$ are shown in Fig. \ref{subfig:phin}.
In Fig. \ref{potential}, we compare the behaviors of  $\tilde{V}(\tilde{y})$  in the three cases of thick branes. Figure ~\ref{potential} shows that the potential $\tilde{V}(\tilde{y})$ in $\text{AdS}_{4}$ brane  opens downwards, different from $\text{dS}_{4}$ brane  where the curve opens upwards. When $\Lambda_{4}\rightarrow0$, $\tilde{V}(\tilde{y})$  in $\text{dS}_{4}$  and $\text{AdS}_{4}$ branes  approach to the values in  $\text{M}_{4}$ brane. The shapes of Lagrange multiplier $\tilde{\lambda}(\tilde{y})$ in terms of  $\tilde{y}=ky$ are plotted in Fig. \ref{lamda}. When $\Lambda_{4}\rightarrow0$, $\tilde{\lambda}(\tilde{y})$ in $\text{dS}_{4}$  and $\text{AdS}_{4}$ branes  also  tend to the values in  $\text{M}_{4}$ brane.

\section{Linear Tensor perturbation} \label{Sec3}
In the case of the tensor perturbation, we suppose that the space-time undergoes a small perturbation $\delta g_{MN}^{(1)}$ on a fixed background $g_{MN}$
\begin{align}
  \tilde{g}_{MN}=g_{MN}+\delta g_{MN}^{(1)},
\end{align}
where $g_{MN}$ represents the five-dimensional Minkowskii, dS, or AdS metric. And the inverse of the perturbed metric will be
\begin{align}
  \tilde{g}^{MN}=g^{MN}+\delta g^{MN(1)}\cdots+\delta g^{MN(n)}+\cdots\,
\end{align}
with the first-order perturbed metric as $\delta g^{MN(1)}=-g^{MP}g^{NQ}\delta g_{PQ}^{(1)}$ and the $n$-order perturbed metric as
$\delta g^{MN(n)}=(-1)^n\delta g^{M(1)}_{P_{1}}\delta g^{P_{1}(1)}_{P_{2}}\cdots\delta g^{P_{n}(1)N}$.
For the above metric perturbations,  the first order perturbations are expressed as follows:
\begin{align}
\delta\Gamma^{L(1)}_{MN}=&\frac{1}{2}g^{LP}\left(\nabla_M\delta g_{NP}^{(1)}+\nabla_N\delta g_{PM}^{(1)}-\nabla_P\delta g_{MN}^{(1)}\right)\  \\
\delta{R}^{L(1)}_{MKN}=&\nabla_{K}\delta\Gamma^{L(1)}_{MN}-\nabla_{N}\delta\Gamma^{L(1)}_{MK}   \\
\delta{R}_{MN}^{(1)}=&\nabla_{K}\delta\Gamma^{K(1)}_{MN}-\nabla_{N}\delta\Gamma^{K(1)}_{MK}
\end{align}
where $\nabla_{N}$ denotes the covariant derivative corresponding to the five-dimensional metric $g_{MN}$.
In general, it is complicated to take into account a full set of fluctuations of the metric around the background  where gravity is coupled to scalars. Fortunately, there is a sector where the metric fluctuations decouple from
the scalars, which is the one associated with the transverse and traceless (TT) part of the
metric fluctuation.  Based on these relations, we will consider the linear tensor perturbation of  flat and bent branes by taking the TT gauge condition.

\subsection{Flat brane}
For the tensor perturbation of flat brane, the perturbed metric is given by
    \begin{eqnarray}
        \label{tensor metric}
        ds^{2}=a^2(y)(\eta_{\mu\nu}+h_{\mu\nu})dx^{\mu}dx^{\nu}+dy^2,
    \end{eqnarray}
where $\eta_{\mu\nu}$ describes Minkowski geometry,  $h_{\mu\nu}$ represents the  tensor perturbation and satisfies TT gauge condition $\eta^{\mu\nu}\partial_\mu h_{\lambda\nu}=0$ and $\eta^{\mu\nu}h_{\mu\nu}=0$.
The linear perturbations of the Ricci tensor and curvature scalar are obtained as
\begin{align}\label{perturbedRMNR}
\delta R_{\mu\nu}^{(1)}&=\frac{1}{2}\left(\partial_{\nu}\partial_{\sigma}h_{\mu}^{\sigma}+\partial_{\mu}\partial_{\sigma}h_{\nu}^{\sigma}-
\square^{(4)}h_{\mu\nu}-\partial_{\mu}\partial_{\nu}h\right) \nonumber \\
&~-\left(3a'^2+aa''\right)h_{\mu\nu}-2aa'h_{\mu\nu}'-\frac{1}{2}a^2h_{\mu\nu}''-\frac{1}{2}aa'\eta_{\mu\nu}h',\nonumber\\
\delta R_{\mu5}^{(1)}&=\frac{1}{2}\left(\partial_\sigma h_\mu^\sigma-\partial_\mu h\right)^{'},\quad
\delta R_{55}^{(1)}=-\frac{1}{2}\left(2a^{-1}a'h'+h''\right),\nonumber\\
\delta R^{(1)}&=\delta\left(g^{MN}R_{MN}\right)=
a^{-2}\left(\partial_{\mu}\partial_{\nu}h^{\mu\nu}-\square^{(4)}h\right)-5a^{-1}a'h'-h'',
\end{align}
where $\square^{(4)}=\eta^{\mu\nu}\partial_{\mu}\partial_{\nu}$ is the four-dimensional d'Alembert operator, and $h=\eta^{\mu\nu}h_{\mu\nu}$.

Under the TT condition, the perturbation of the $\mu\nu$ components of the Einstein tensor reads
    \begin{eqnarray}
    \label{perturbation Gmn}
        \delta G_{\mu\nu}^{(1)}=-\frac{1}{2}\Box^{(4)}h_{\mu\nu}+(3a'^2+3aa'')h_{\mu\nu}-2aa'h'_{\mu\nu}-\frac{1}{2}a^2 h''_{\mu\nu},
    \end{eqnarray}
where the four-dimensional d'Alembertian is defined as $\Box^{(4)}\equiv\eta_{\mu\nu}\partial_{\mu}\partial_{\nu}$.
Using Eqs. (\ref{eom21}) and (\ref{perturbation Gmn}), the perturbation equation reads
    \begin{eqnarray}
        -\frac{1}{2}\Box^{(4)}h_{\mu\nu}
        -2aa'h'_{\mu\nu}-\frac{1}{2}a^2 h''_{\mu\nu}=0.\label{equationh}
    \end{eqnarray}
By imposing a coordinate transformation,  $ dz=\frac{1}{a(z)}dy $ and a rescaling on
$h_{\mu\nu}=a(z)^{-3/2}\tilde{h}_{\mu\nu}$, the perturbation equation (\ref{equationh}) can be calculated as:
    \begin{eqnarray}
        \Box^{(4)}\tilde{h}_{\mu\nu}+\partial^2_{z}\tilde{h}_{\mu\nu}
        -\frac{\partial^2_{z}a^{\frac{3}{2}}(z)}{a^{\frac{3}{2}}(z)}\tilde{h}_{\mu\nu}=0.
    \end{eqnarray}
Considering the Kaluza-Klein (KK) decomposition $\tilde{h}_{\mu\nu}=\epsilon_{\mu\nu}(x^\gamma) \text{e}^{ip_{\lambda}x^{\lambda}}H(z)$ with $p^{2}=-m^{2}$, where the polarization tensor $\epsilon_{\mu\nu}$ satisfies the TT condition $\eta^{\mu\nu}\partial_\mu \epsilon_{\lambda\nu}=0$ and $\eta^{\mu\nu}\epsilon_{\mu\nu}=0$, we obtain the Schr\"{o}dinger-like equation for $H(z)$:
    \begin{eqnarray}
        \label{eq tensor}
        \left[-\partial^2_{z}+V_\text{eff}(z)\right]H(z)=m^2 H(z),
    \end{eqnarray}
where $m$ is the mass of the Kaluza-Klein (KK) mode, and  the effective potential $V_\text{eff}(z)$ is given by \cite{Afonso:2006gi}
    \begin{eqnarray}
        V_\text{eff}(z)=\frac{\partial^2_{z}a^{\frac{3}{2}}(z)}{a^{\frac{3}{2}}(z)}=\left(\partial_{z} \ln a^{\frac{3}{2}}(z)\right)^2+\partial_{z}\left(\partial_{z} \ln a^{\frac{3}{2}}(z)\right).\label{effective potential}
    \end{eqnarray}
This equation can be factorized as
\begin{eqnarray}
-\bigg[\partial_{z}+ \partial_{z} \ln a^{\frac{3}{2}}(z)\bigg]\bigg[\partial_{z}- \partial_{z} \ln a^{\frac{3}{2}}(z)\bigg] H(z)=m^2 H(z), \label{eq tensor1}
\end{eqnarray}  and this structure  ensures that the eigenvalues are non-negative, which means that the brane is stable against the tensor perturbation. Since the potential vanishes for large $z$, this is the only bound state, namely, the massless zero mode ($m=0$),
\begin{eqnarray}
        H_0 (z(y))\propto a^{\frac{3}{2}}(z(y))=\bigg[\text{sech}(k (y-b))+\text{sech}(k y)+\text{sech}(k (y+b))\bigg]^{3/2}.
    \end{eqnarray}
 To localize the gravity zero-mode, $H_{0}(z)$ should obey the normalization
condition $\int_{-\infty}^{+\infty}\left|H_0 (z)\right|^2 \mathrm{d}z<\infty $. It can be normalized if
\begin{eqnarray}\label{normalization1}
  &&\int_{-\infty}^{+\infty}\left|H_0 (z)\right|^2 \mathrm{d}z=\int_{-\infty}^{+\infty}\left|H_0 (z(y))\right|^2 a(y)^{-1}\,\mathrm{d}y\nonumber \\
  &=&\int_{-\infty}^{+\infty}\bigg[\text{sech}(k (y-b))+\text{sech}(k y)+\text{sech}(k (y+b))\bigg]^{2}dy<\infty,
\end{eqnarray}
which is finite when $k>0$; in other words, the normalized zero-mode can be achieved for $k>0$ such that the observable four-dimensional gravity is recovered on the brane. The behaviors of the dimensionless effective potential and zero-mode  in terms of $\tilde{y}$ are shown in  Fig.~\ref{effectivepotential2}.  The effective potentials have a well with a negative minimum inside the brane and satisfy
$V_\text{eff}(\tilde{y}\rightarrow\pm\infty)\rightarrow 0$ when far from the brane. As the parameter $\tilde{b}$ increases,  the volcano-like potential gradually changes to a multi-well potential, and at last splits into three-well potential; meanwhile, the wave function of the graviton zero-mode also splits.
\begin{figure}[!htb]
    \begin{center}
    \subfigure[$\tilde{b}=0.5$]{\label{fig model 1}
        \includegraphics[width=2.0in]{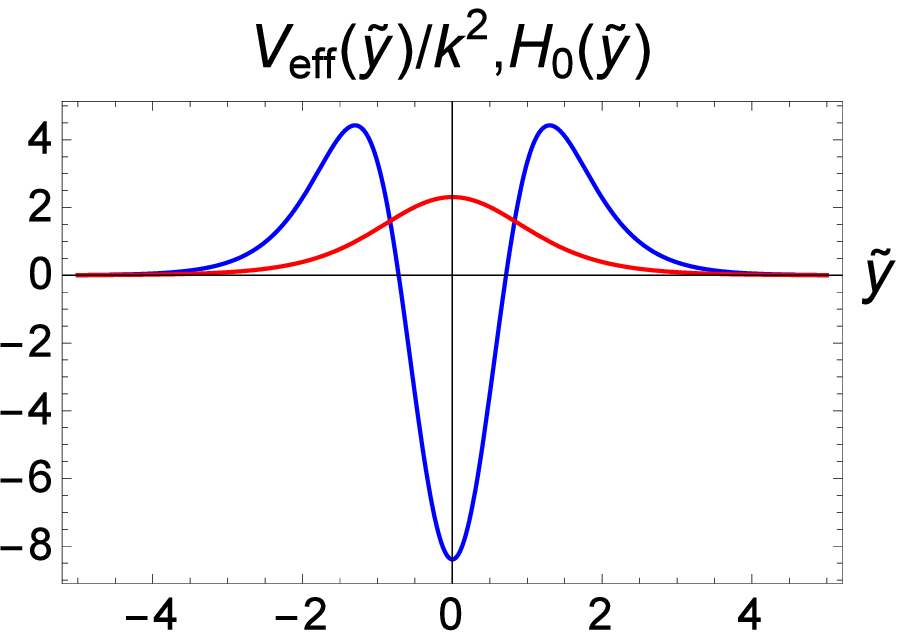}}
     \subfigure[$\tilde{b}=3$]{\label{fig model 2}
     \includegraphics[width=2.0in]{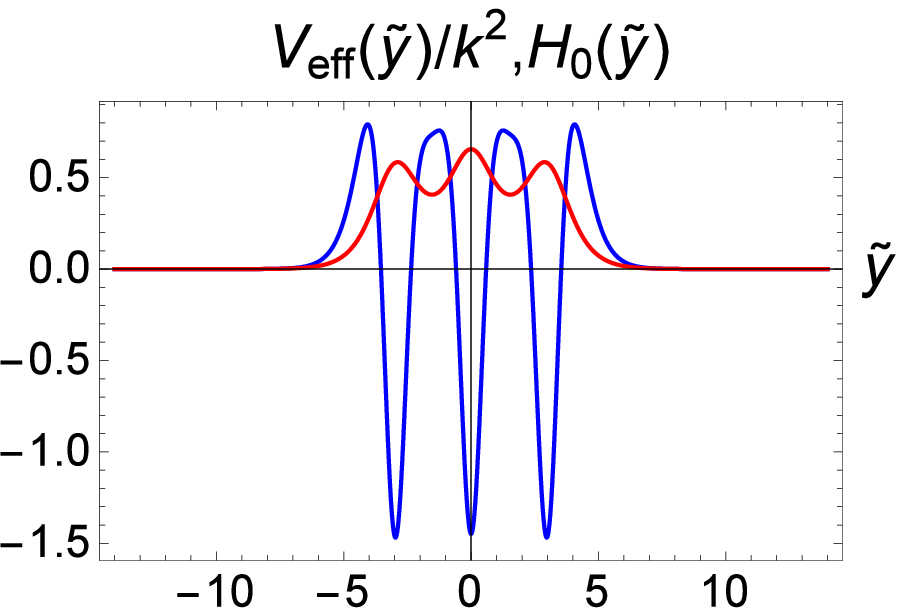}}
     \subfigure[$\tilde{b}=8$]{\label{fig model 2}
     \includegraphics[width=2.0in]{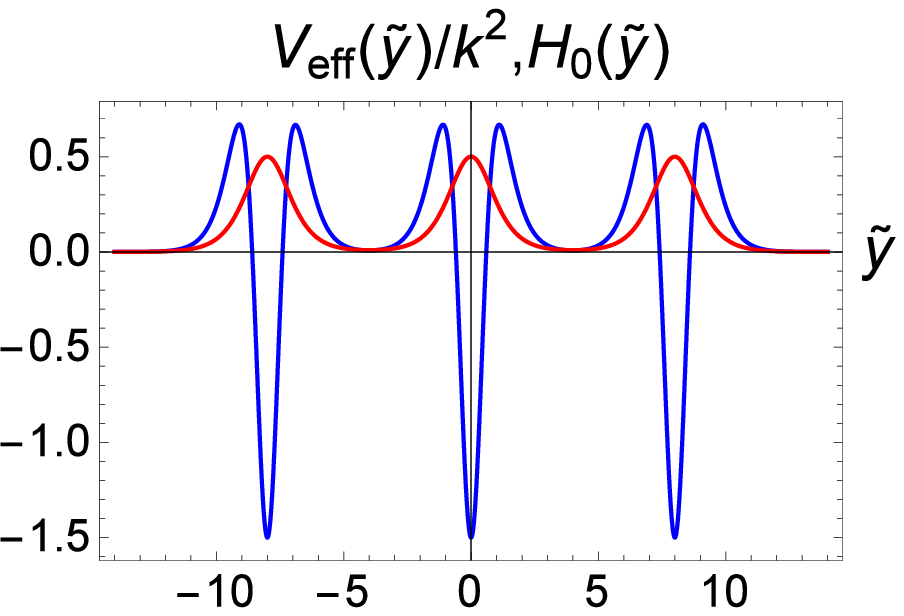}}
    \end{center}
    \caption{The influence of $\tilde{b}$ on the  dimensionless effective potential $V_\text{eff}(\tilde{y})/k^{2}$ (blue lines) and the zero-mode $H_{0}(\tilde{y})$ (red lines)  of the tensor perturbation for $\text{M}_{4}$ brane. }
    \label{effectivepotential2}
    \end{figure}

\subsection{Bent branes}
We now turn  to the case of $\Lambda_4\neq0$ and  consider the following perturbed metric,
\begin{eqnarray}
        \label{tensor metric}
        ds^{2}=a^2(y)(\hat{g}_{\mu\nu}+h_{\mu\nu})dx^{\mu}dx^{\nu}+dy^2,
    \end{eqnarray}
 where the four-dimensional metric is decomposed into a small perturbation $h_{\mu\nu}$ around a curved space-time $a^2(y)\hat{g}_{\mu\nu}$.
 By imposing a coordinate transformation $ dz=\frac{1}{a(z)}dy $, we write the bulk metric in the form
\begin{eqnarray}
        \label{tensor metric}
        ds^{2}=a^2(z)\bigg[(\hat{g}_{\mu\nu}+h_{\mu\nu})dx^{\mu}dx^{\nu}+dz^2\bigg].
    \end{eqnarray}
The interesting investigations of the tensor perturbation have been already appeared in
Refs. \cite{Kobayashi2002,RandallJHEP2001} and in references therein. Then, under the TT gauge condition $h^{\mu}_{\mu}=\hat{\nabla}^{\mu}h_{\mu\nu}=0$, the equation for the perturbation $h_{\mu\nu}$ takes the following form:
\begin{eqnarray}\label{dynamics_eq}
\left[\partial_{z}^{2} + 3\bigg(\partial_{z}a(z)/{a(z)}\bigg)\partial_{z}+
 \hat{g}^{\alpha\beta}\hat{\nabla}_{\alpha}\hat{\nabla}_{\beta}
   -\frac{9}{4}\Lambda_{4}\right]h_{\mu\nu}(x,z)=0,
\end{eqnarray}
where $\hat{\nabla}_{\alpha}$ denotes the covariant derivative with
respect to $\hat{g}_{\mu\nu}$.
By performing the KK decomposition $h_{\mu\nu}(x,z) =a(z)^{-3/2}\epsilon_{\mu\nu}(x)H(z)$ with $\epsilon_{\mu\nu}(x)$ satisfying the TT condition, we separate the perturbed  equation (\ref{dynamics_eq}) into  the four-dimensional and extra-dimensional part. Here  we can get two  equations, i.e.,  $\hat{g}^{\alpha\beta}\hat{\nabla}_{\alpha}\hat{\nabla}_{\beta}\epsilon_{\mu\nu}(x)=m^{2}\epsilon_{\mu\nu}(x)$ for the four-dimensional part, and the Schr\"{o}dinger-like equation for the extra-dimensional sector
\begin{eqnarray}\label{Schrodinger_eq2}
\left[-\partial_{z}^{2}+V_\text{eff}(z)\right]H(z)=m^{2}H(z).
\end{eqnarray}
Here $m$ is the mass of the KK mode, and the effective potential is derived as
\begin{eqnarray}\label{VQM}
V_\text{eff}(z)=-\frac{9}{4}\Lambda_{4}+\frac{3}{4}\frac{[\partial_{z}a(z)]^{2}}{a(z)^{2}}+\frac{3}{2}\frac{\partial_{z}\partial_{z}a(z)}{a(z)},
\end{eqnarray}
for $dS_{4}$ geometry.  The effective potential can also be  transformed  in terms of $y$ coordinate,
\begin{eqnarray}\label{VQM}
V_\text{eff}(z(y))=-\frac{9}{4}\Lambda_{4}+\frac94\left[\partial_y a(y)\right]^2+\frac32a(y)\partial_{y,y}a(y),
\end{eqnarray}
At the boundaries of the brane, the potential $V_\text{eff}(z)$ tends to be negative   for  $\Lambda_{4}>0$. For the term $-\frac{9}{4}\Lambda_{4}< 0$, this equation can not be factorized. This result indicates that the tensor perturbation of $\text{dS}_{4}$ brane would not occur stably. For
 $\Lambda_{4}<0$, we get  $\text{AdS}_{4}$  geometry, and the potential $V_\text{eff}(z)$ tends to be positive  at $z\rightarrow\pm\infty$.
Eq. (\ref{Schrodinger_eq2}) can  be written as a factorizable  equation, $ \mathcal{K}\mathcal{K}^{\dagger}H(z) =m^2 H(z)$
 with \begin{eqnarray}
 \mathcal{K} = -\partial_{z}- \partial_{z} \ln a^{\frac{3}{2}}(z)+\sqrt{-9\Lambda_{4}/4},
 \mathcal{K}^{\dag} = \partial_{z}- \partial_{z} \ln a^{\frac{3}{2}}(z)+\sqrt{-9\Lambda_{4}/4},
 \end{eqnarray}
which ensures the stability of the tensor perturbation.

\section{Massive resonant modes}
For the volcano-like effective potentials, the tensor perturbation has  zero-mode and may also have resonant modes.
A further investigation of the  metastable modes is necessary. Because the integral $z=\int\frac{1}{a(y)}dy$ is difficult, we can
not obtain the analytical expressions of the warp factor $a(z)$ and the effective potential
$V_\text{eff}(z)$. To solve the Schr\"{o}dinger-like equation  (\ref{eq tensor1}) for $H(z)$ numerically, we decompose
$H(z)$ into an even parity mode  and an odd parity mode, which are set to satisfy
the following boundary conditions
 \begin{eqnarray}
        \label{condition even}
        H_\text{even} (0)=1,~~~~~~~~\partial_z H_\text{even} (0)=0;\\
        \label{condition odd}
        H_\text{odd} (0)=0,~~~~~~~~\partial_z H_\text{odd} (0)=1.
    \end{eqnarray}
To find the massive resonant states,  we use the numerical method given in Refs.~\cite{Liu2009a,Liu2009c,Du2013,Tan:2020sys},  where a relative probability   was proposed:
\begin{eqnarray}
P=\frac{\int_{-z_c}^{z_c}|H_{n}(z)|^2 \text{d}z}
     {\int_{-z_{max}}^{z_{max}}|H_{n}(z)|^2 \text{d}z}.
\end{eqnarray}
{Here} $2z_c$ is about the width of the thick brane and $z_{max}$ is set to $z_{max}=10z_c$.

\subsection{Flat brane}
When the wave functions are either even-parity or odd-parity, the Schr\"{o}dinger-like equation can be solved numerically. The dimensionless effective potential $V_\text{eff}(\tilde{z})/k^{2}$ is expressed in terms of $\tilde{z}=kz$. Figure \ref{effectivepotential2} shows the influence of $\tilde{b}$ on the  effective potential and the resonant  modes of  gravity. The relative probability  $P$ as a function of $m^{2}$ is obtained, and  only the peak which satisfies $P>0.1$ represents a resonance mode.
  \begin{figure}[!htb]
    \begin{center}
    \subfigure[$\tilde{b}=0.5$]{\label{fig model 1}
        \includegraphics[width=2.0in]{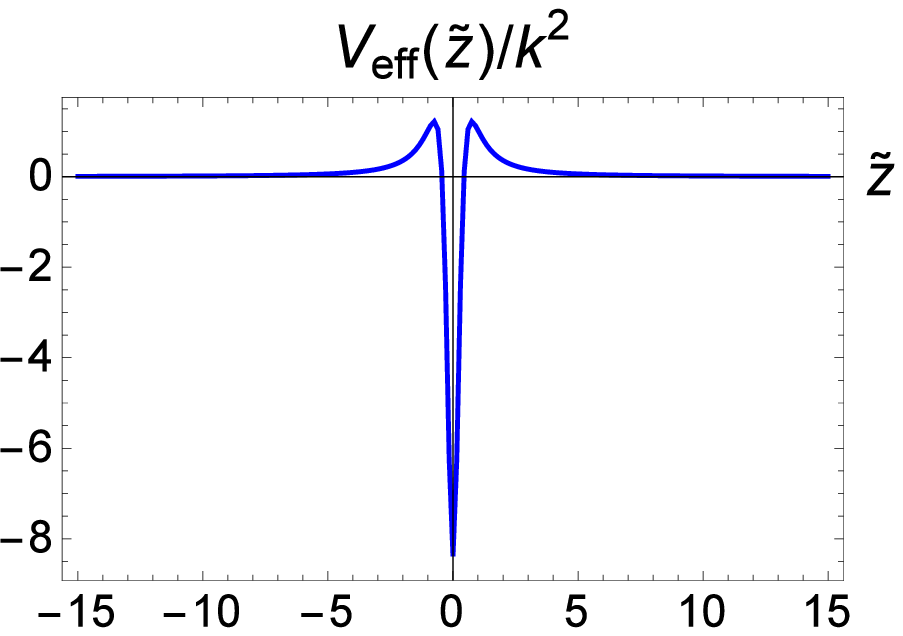}}
     \subfigure[$\tilde{b}=3$]{\label{fig model 2}
     \includegraphics[width=2.0in]{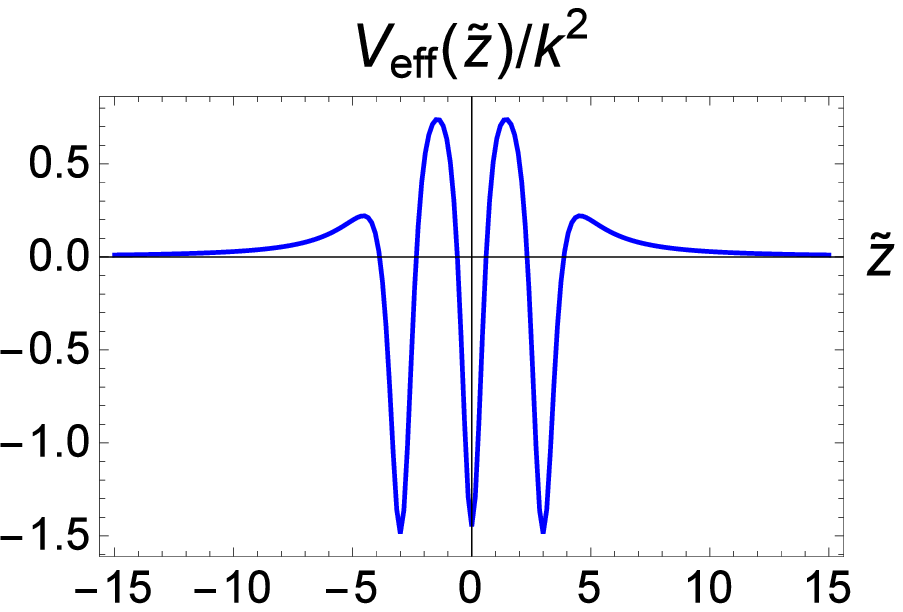}}
     \subfigure[$\tilde{b}=8$]{\label{fig model 2}
     \includegraphics[width=2.0in]{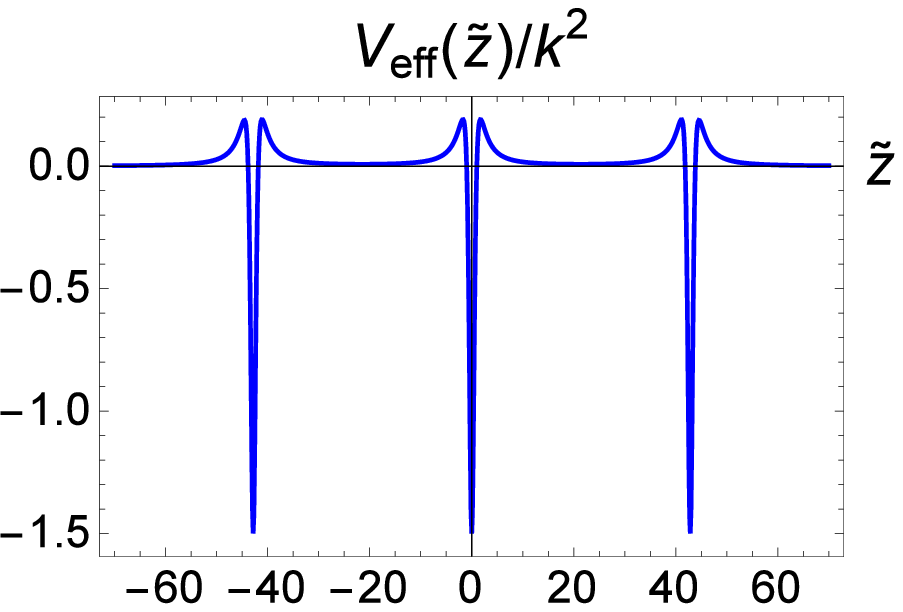}}
     \subfigure[$\tilde{b}=0.5$]{\label{subfig:relative probability1}
        \includegraphics[width=2.0in]{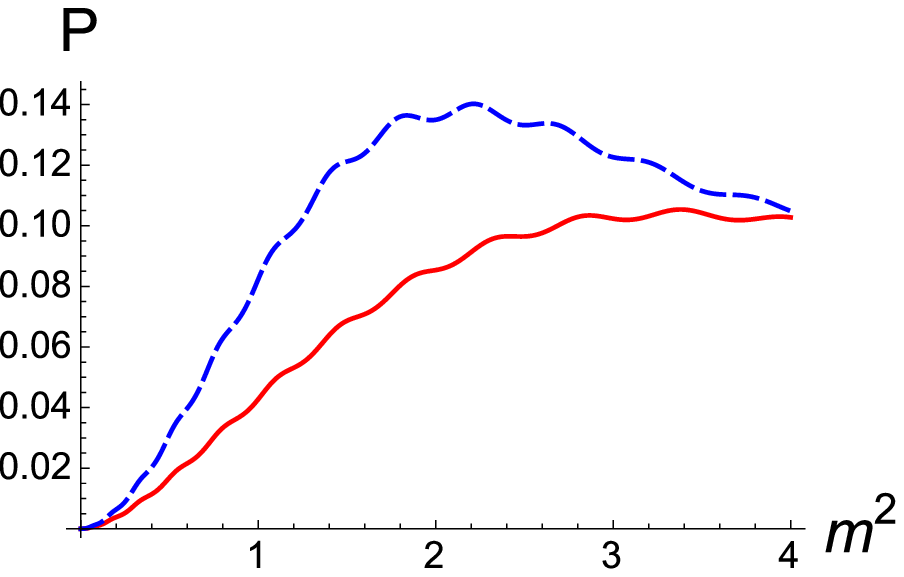}}
     \subfigure[$\tilde{b}=3$]{\label{subfig:relative probability2}
     \includegraphics[width=2.0in]{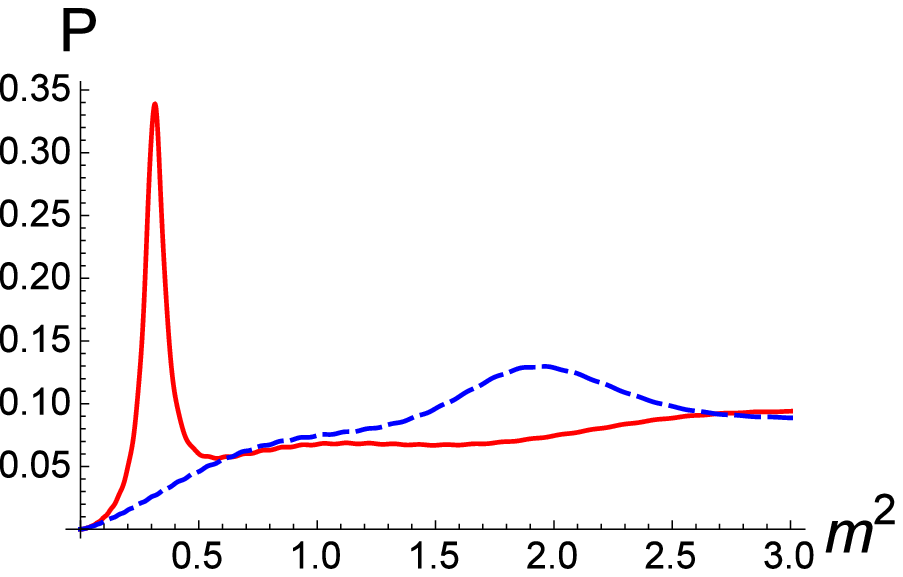}}
     \subfigure[$\tilde{b}=8$]{\label{subfig:relative probability3}
     \includegraphics[width=2.0in]{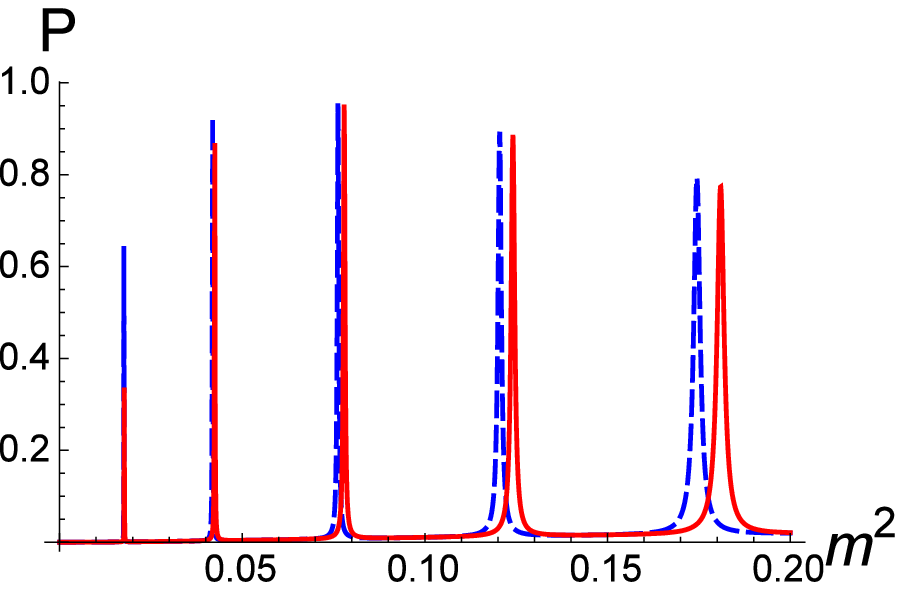}}
    \end{center}
    \caption{The influence of the  parameter $\tilde{b}$ on the effective potential  $V_\text{eff}(\tilde{z})/k^{2}$ and the probabilities $P$  for the odd-parity (blue dashed lines) and even-parity (red lines) massive KK modes. }
    \label{effectivepotential2}
    \end{figure}

\begin{figure}[!htb]
    \begin{center}
    \subfigure[ $\tilde{b}=3$ ]{\label{subfig:effectivepotential21}
    \includegraphics[width=2.0in]{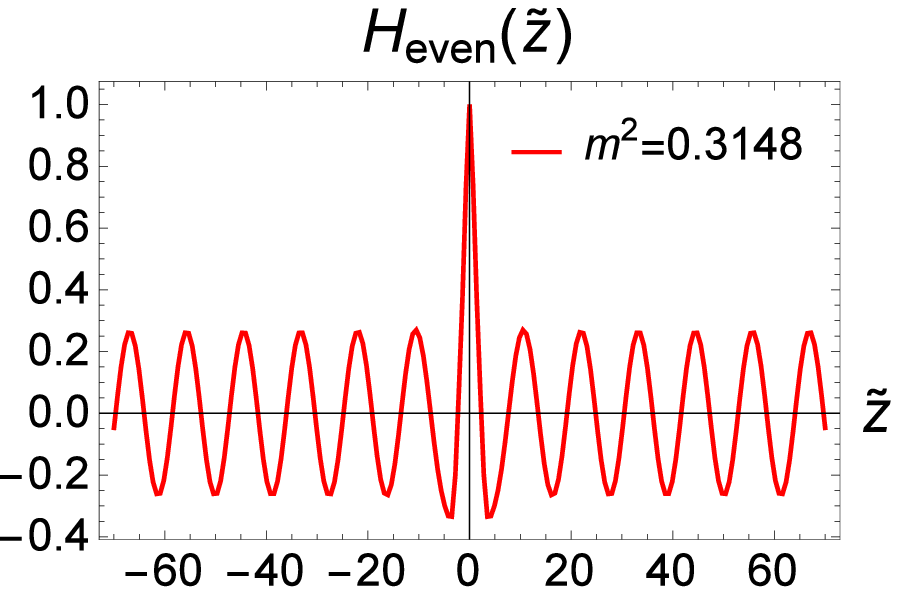}~~~~~~~
    \includegraphics[width=2.0in]{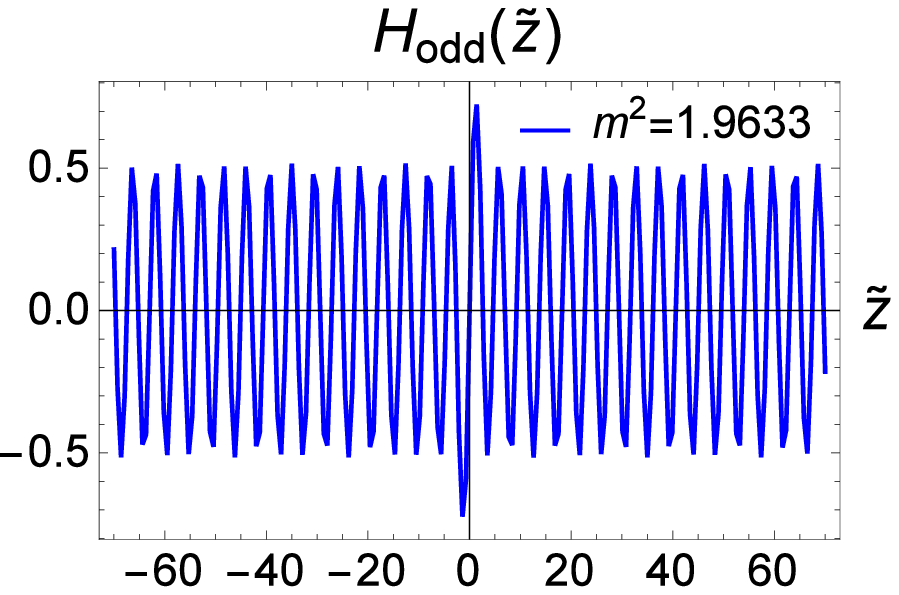}}\\
    \subfigure[ $\tilde{b}=8$]{\label{subfig:effectivepotential21}
    \includegraphics[width=2.0in]{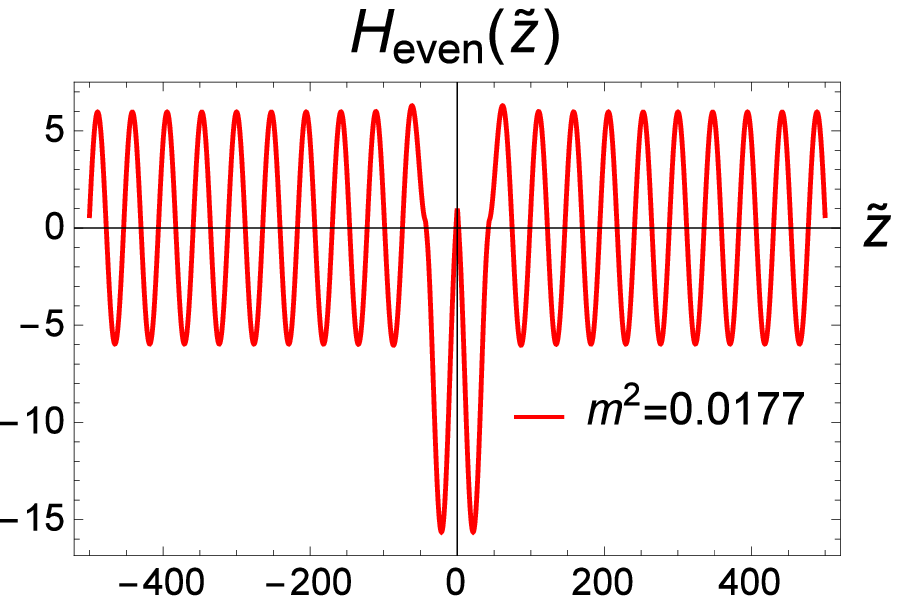}~~~~~~~
   \includegraphics[width=2.0in]{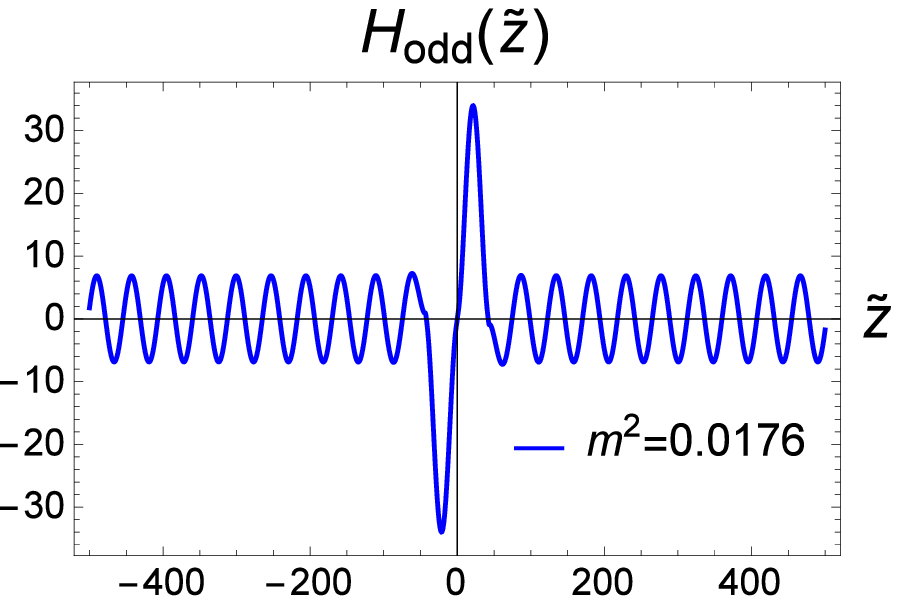}}
    \end{center}
    \caption{The wave functions for the first  even-parity  and  odd-parity  modes  for the flat brane ($\tilde{\Lambda}_{4}=0$) with  $\tilde{b}=3$ and $\tilde{b}=8$.}
    \label{wavefunction4}
    \end{figure}

From Fig.~\ref{effectivepotential2}, we see that, with the increasing of the parameter  $\tilde{b}$ the effective  potential splits from a single-well into a three-well potential,   which indicates there are
more resonant KK modes for larger $\tilde{b}$, and this can be confirmed by Figs. \ref{subfig:relative probability1},\;\ref{subfig:relative probability2},\; \ref{subfig:relative probability3}. In Fig.~\ref{subfig:relative probability1}, there are no peaks of the relative probability, which means there  does not exist resonant mode. In Fig.~\ref{subfig:relative probability2}, there is just one peak of the relative probability corresponding to the even-parity  or the odd-parity wave function, and its wave functions  with mass square $m^2=0.3148$ and $m^2=1.9633$ are plotted in Fig.~\ref{wavefunction4}, which shows that the resonance is indeed quasi-localized on the sub-brane. In Fig. \ref{subfig:relative probability3},  we find there are five peaks corresponding to the even-parity  or the odd-parity resonant modes  which satisfy $P>0.1$, and the corresponding wave functions of the first even-parity and odd-parity modes with mass square $m^2=0.0177$ and $m^2=0.0176$ are plotted in Fig. \ref{wavefunction4}. The numerical results of the mass spectrum, relative probability, full width at half maximum (FWHM) and lifetime of the
resonance with $\tilde{b}=3,8$ are listed in Tab. \ref{Table2}.
The resonances having a large lifetime can be quasi-localized on the brane for a long time. Note that the first even and odd resonance modes are not
degenerate.

\begin{table*}[htb]
\begin{center}
\begin{tabular}{||c|c|c|c|c|c|c|c||}
\hline
$\tilde{b}$ & $\tilde{\Lambda}_{4}$ &$n$ &parity & $m_{n}^2$ &  $P$ & $\Gamma$ & $\tau$    \\
\hline \hline
&0&1 &even & $0.1988$ &$0.372419$ & $0.00810986$ & $123.307$ \\ \cline{2-8}
\raisebox{2.3ex}[0pt]{3}
&0&2 &odd & $1.8477$ &$0.144147$ & $1.15022$ & $0.8694$ \\ \cline{2-8}
\hline\hline
&0&1 & odd & $0.0176$ &$0.644363$ & $0.00112908$ & $885.681$ \\ \cline{2-8}
&0&2 & even & $0.0177$ &$0.336324$ & $0.00112589$ & $888.186$ \\ \cline{2-8}
&0&3 & odd & $0.0419$ &$0.923572$ & $0.000732362$ & $1365.44$ \\ \cline{2-8}
&0&4 & even & $0.0424$ &$0.850191$ & $0.000727779$ & $1374.04$ \\ \cline{2-8}
&0&5 & odd & $0.0763$ &$0.933456$ & $0.000905549$ & $1104.30$ \\ \cline{2-8}
&0&6 & even & $0.0779$ &$0.94479$ & $ 0.00089571$ & $1116.43$  \\ \cline{2-8}
\raisebox{2.3ex}[0pt]{8}
&0&7 & odd & $0.1205$ &$0.909459$ & $0.00172875$ & $578.454$ \\ \cline{2-8}
&0&8 & even & $0.1241$ &$0.89555$ & $0.00141991$ & $587.129$ \\ \cline{2-8}
&0&9 & odd & $0.1745$ &$0.792877$ & $0.00215697$ & $278.884$ \\ \cline{2-8}
&0&10  & even & $0.1808$ &$0.755761$ & $0.00358572$ & $404.900$ \\ \cline{2-8}
\hline
\end{tabular}\\
\caption{The influence of the parameter $\tilde{b}$ on the mass spectrum $m_{n}$, the relative probability $P$, the width of mass $\Gamma$, and the lifetime $\tau$ of the KK resonances for the flat brane ($\tilde{\Lambda}_{4}=0$).}
\label{Table2}
\end{center}
\end{table*}

\subsection{Bent branes}
Because the tensor perturbation of $\text{dS}_{4}$ brane would not be stable, we now consider  $\text{AdS}_{4}$ brane,  which include two parameters $\tilde{b}$ and $\tilde{\Lambda}_{4}$. Thus, the resonances are more involved and should be discussed specifically.
\begin{figure}[!htb]
    \begin{center}
    \subfigure[$V_\text{eff}(\tilde{z})/k^{2}$]{\label{subfig:effectivepotential21}
        \includegraphics[width=2.0in]{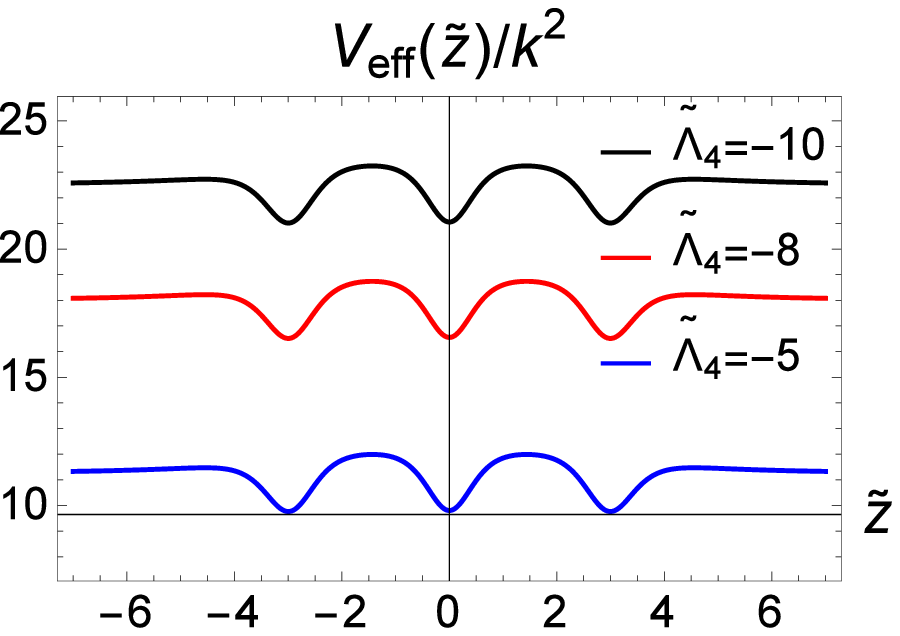}}
     \subfigure[$\tilde{\Lambda}_{4}=-5$]{\label{subfig:effectivepotential22}
     \includegraphics[width=2.0in]{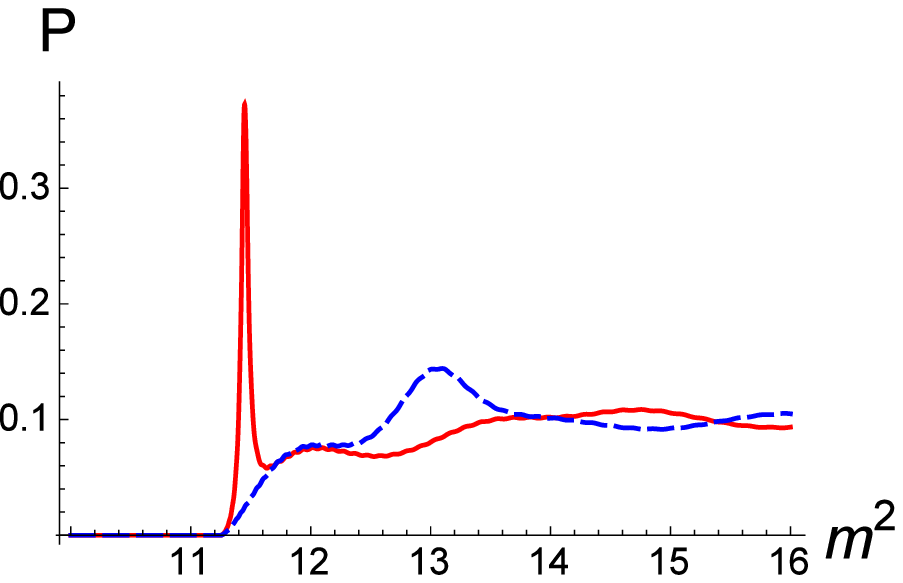}}\\
      \subfigure[$\tilde{\Lambda}_{4}=-8$]{\label{subfig:effectivepotential23}
     \includegraphics[width=2.0in]{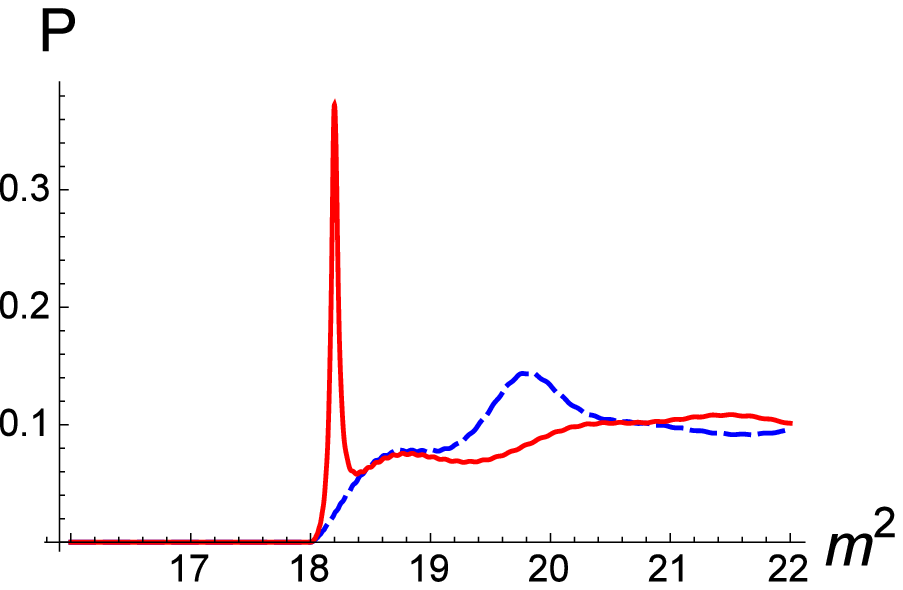}}
     \subfigure[$\tilde{\Lambda}_{4}=-10$]{\label{subfig:effectivepotential24}
     \includegraphics[width=2.0in]{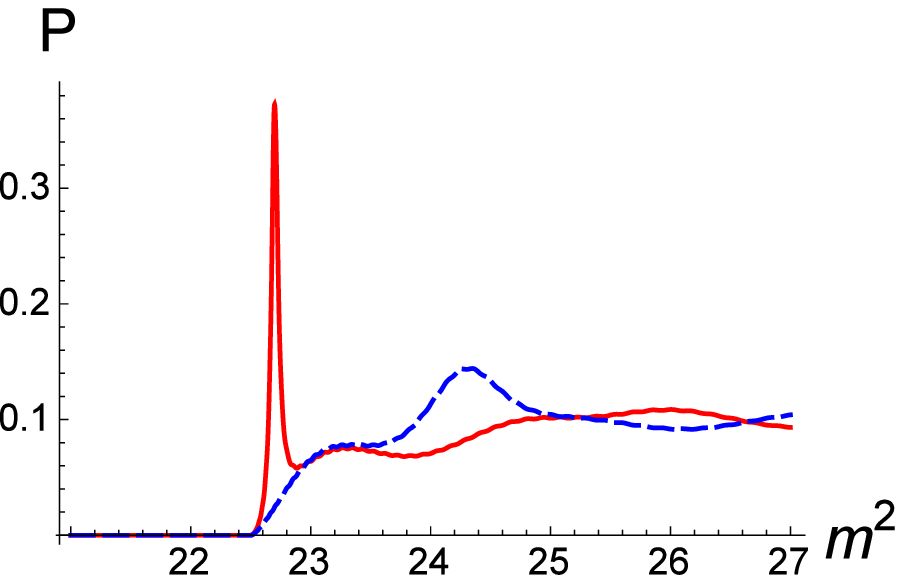}}
    \end{center}
    \caption{The influence of the  parameter $\tilde{\Lambda}_{4}$ on the effective potential  and the probabilities  for
both the odd-parity (blue dashed lines) and even-parity (red
lines) massive KK modes with a fixed parameter $\tilde{b}=3$. }
    \label{fig:effectivepotential1}
\end{figure}

\begin{figure}[!htb]
    \begin{center}
    \subfigure[ odd-parity ]{\label{subfig:effectivepotential21}
    \includegraphics[width=2.0in]{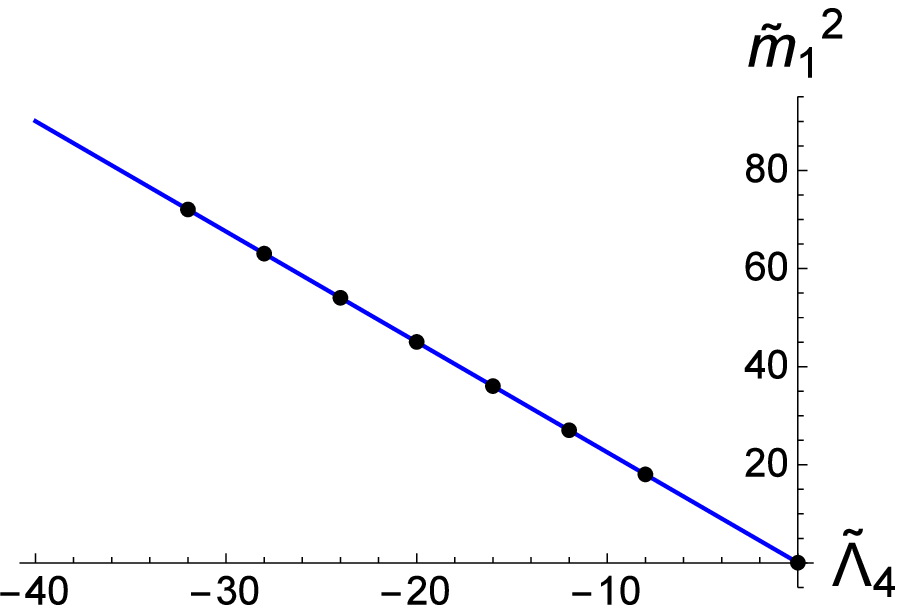}}
    \subfigure[ even-parity]{\label{subfig:effectivepotential21}
   \includegraphics[width=2.0in]{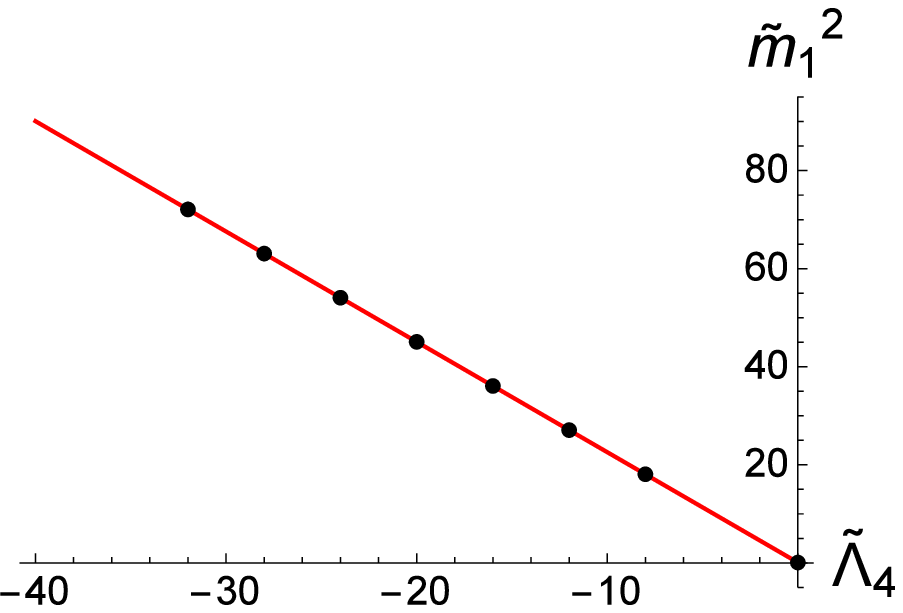}}
    \end{center}
    \caption{The influence of $\tilde{\Lambda}_{4}$ on the masses of  the first  even-parity  and  odd-parity  modes   with $\tilde{b}=8$. The black dots are numerical results, the  solid lines  are  the fit functions for the first even-parity (red line) and odd-parity (blue line) modes.  }
    \label{wavefunction5}
    \end{figure}

\begin{figure}[!htb]
    \begin{center}
    \subfigure[ $\tilde{\Lambda}_{4}=-8$ ]{\label{subfig:effectivepotential21}
    \includegraphics[width=2.0in]{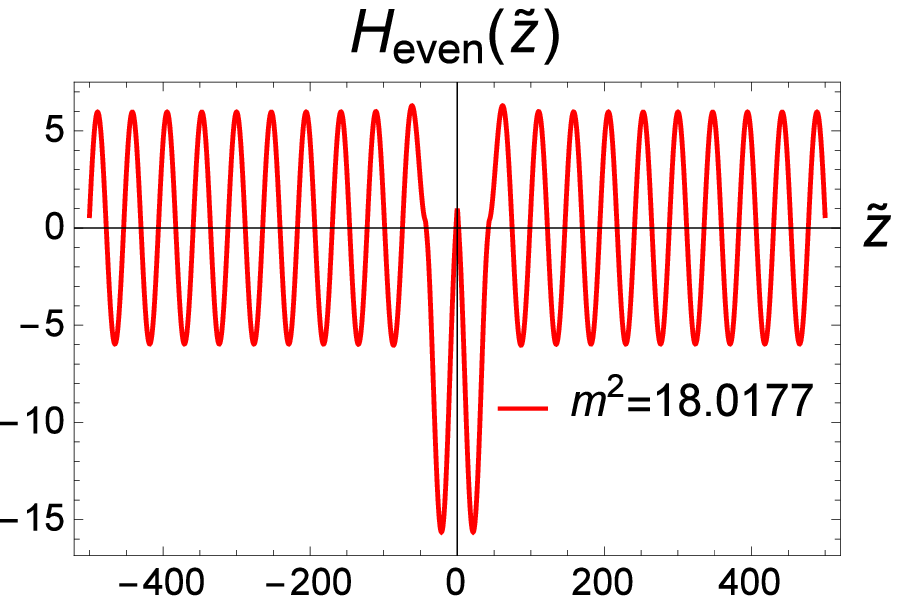}~~~~~~~
    \includegraphics[width=2.0in]{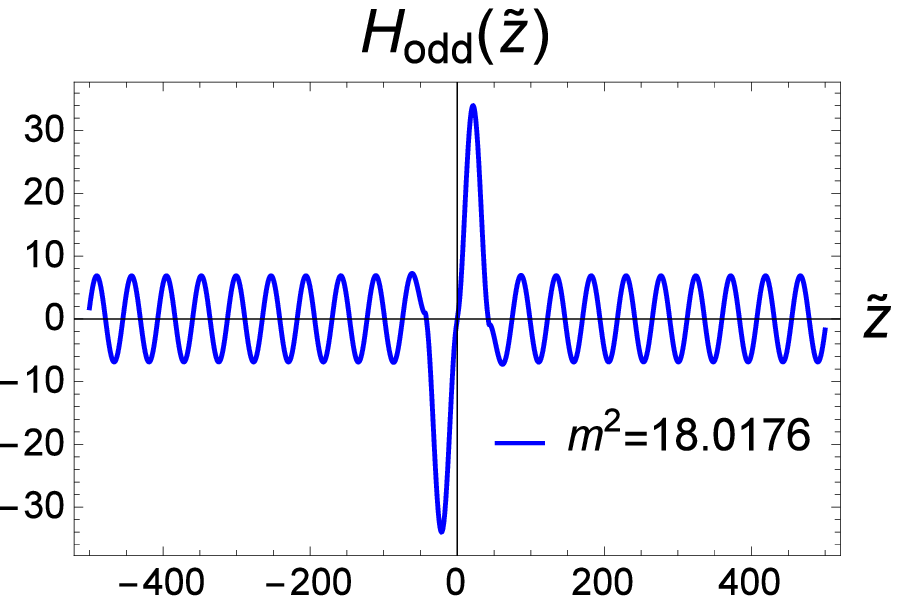}}\\
    \subfigure[$\tilde{\Lambda}_{4}=-12$]{\label{subfig:effectivepotential21}
    \includegraphics[width=2.0in]{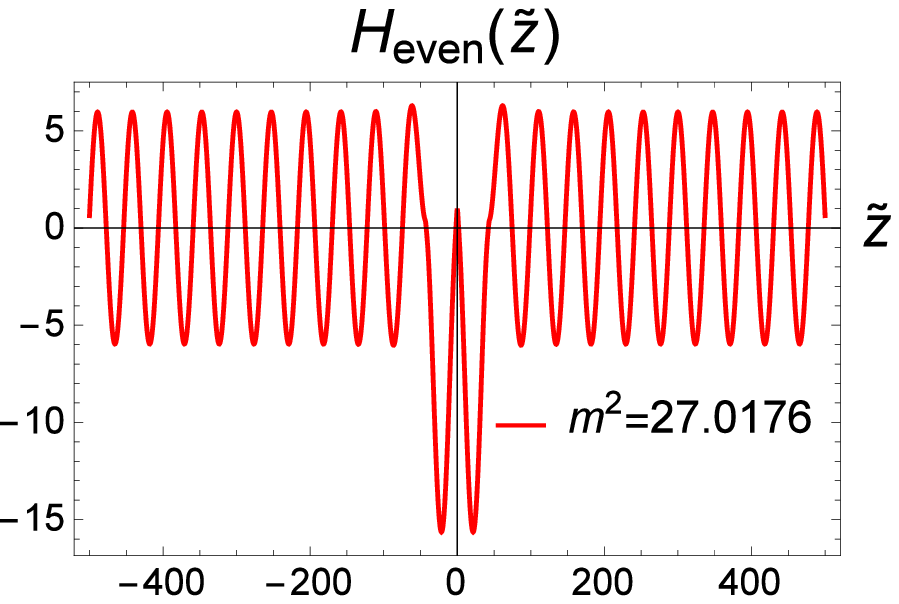}~~~~~~~
   \includegraphics[width=2.0in]{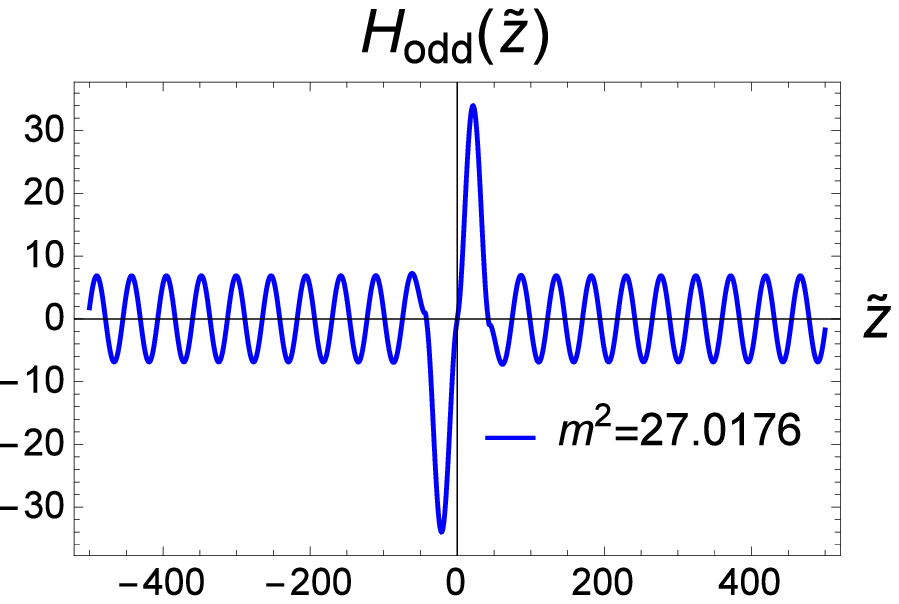}}
    \end{center}
    \caption{The wave functions for the first  even-parity  and first odd-parity  modes   with $\tilde{b}=8$.}
    \label{wavefunction3}
    \end{figure}

The effects of the  parameters $\tilde{\Lambda}_{4}$  on the effective potentials and relative probability are shown in Fig. \ref{fig:effectivepotential1}.  The effective potentials have a  multi-well with a minimum inside the brane and satisfy $\tilde{V}_\text{eff}(\tilde{z})\rightarrow -\frac{2}{3}\tilde{\Lambda}_{4}$ when $\tilde{z}\rightarrow\pm\infty$. The effective potential does not change its shape but shift up and down  with $\tilde{\Lambda}_{4}$. Therefore, the number of the resonances,  the relative probability $P$, the width of mass $\Gamma$, and the lifetime $\tau$ of the KK resonances do not change with $\tilde{\Lambda}_{4}$ for a fixed $\tilde{b}$. Only the mass spectrum changes with $\tilde{\Lambda}_{4}$.  The specific values of masses of the resonances with different values of $\tilde{\Lambda}_{4}$  (including $\tilde{\Lambda}_{4}=0$) are listed in Tab. \ref{tableA}, from which we obtain that  all of the  masses of the resonances depend linearly on the parameter $\tilde{\Lambda}_{4}$. For instance,  the relations for the masses of the first  even-parity  and  odd-parity  modes   with $\tilde{b}=8$  can be expressed as
\begin{table}
\centering
\begin{tabular}{|c|c|c||c|c|c||c|c|c|}\hline
\multicolumn{3}{|c||}{$\tilde{\Lambda}_{4}=0$}&
\multicolumn{3}{|c||}{$\tilde{\Lambda}_{4}=-8$}&\multicolumn{3}{|c|}{$\tilde{\Lambda}_{4}=-12$}
\\\hline
\multicolumn{1}{|c|}{$\tilde{b}=3$}&\multicolumn{1}{|c|}{$\tilde{b}=7$}&
\multicolumn{1}{|c||}{$\tilde{b}=8$}&
\multicolumn{1}{|c|}{$\tilde{b}=3$}&\multicolumn{1}{|c|}{$\tilde{b}=7$}&
\multicolumn{1}{|c||}{$\tilde{b}=8$}&
\multicolumn{1}{|c|}{$\tilde{b}=3$}&
\multicolumn{1}{|c|}{$\tilde{b}=7$}&\multicolumn{1}{|c|}{$\tilde{b}=8$}
\\\hline
$0.1988$&$0.047$&$0.0176$&
$18.1988$&$18.047$&$18.0176$&$27.1988$&$27.047$&$27.0176$
\\\hline
$1.8477$&$0.0481$&$0.0177$&
$19.8477$&$18.0481$&$18.0177$&28.8477&$27.0481$&$27.0177$
\\\hline
-&$0.1104$&$0.0419$&
-&$18.1104$&$18.0419$&-&$27.1104$&$27.0419$
\\\hline
-&$0.1156$&$0.0424$&
-&$18.1156$&$18.0424$&-&$27.1156$&$27.0424$
\\\hline
-&$0.199$&$0.0763$&
-&$18.199$&$18.0763$&-&$27.199$&$27.0763$
\\\hline
-&$0.2122$&$0.0779$&
-&$18.2122$&$18.0779$&-&$27.2122$&$27.0779$
\\\hline
-&$0.3123$&$0.1205$&
-&18.3123&$18.1205$&-&$27.3123$&$27.1205$
\\\hline
-&$0.3375$&$0.1241$&
-&18.3375&$18.1241$&-&27.3375&$27.1241$
\\\hline
-&-&$0.1745$&
-&-&$18.1745$&-&-&27.1745
\\\hline
-&-&$0.1808$&
-&-&$18.1808$&-&-&27.1808
\\\hline
\end{tabular}
\caption{The influence of the parameter $\tilde{\Lambda}_{4}$ on  the masses $m_{n}^{2}$ of resonances for the KK modes.}
\label{tableA}
\end{table}
\begin{eqnarray}
\tilde{m}_{1\text{odd}}^{2}&=&0.0176-\frac{9}{4}\tilde{\Lambda}_{4},  \\
\tilde{m}_{1\text{even}}^{2}&=&0.0177-\frac{9}{4}\tilde{\Lambda}_{4}.
\end{eqnarray}
We plot the fit functions for masses of  the first even-parity and odd-parity modes  with different $\tilde{\Lambda}_{4}$ in Fig. \ref{wavefunction5}.
The wave functions for the first  even-parity  and  odd-parity  modes   with different $\tilde{\Lambda}_{4}$ are plotted in Fig. \ref{wavefunction3}, which shows that the wave function does not alter  with $\tilde{\Lambda}_{4}$ when $\tilde{b}$ is fixed.

For $\text{AdS}_{4}$ brane, the potential well also becomes splitting, and the number of resonant modes increases with the parameter $\tilde{b}$. The influence of the  parameter $\tilde{b}$ on the effective potential  $V_\text{eff}(\tilde{z})/k^{2}$ and the probabilities $P$ is similar to flat brane, so we do not discuss it repeatedly. Here  we list the numerical results for mass spectrum, the relative probability, the width of mass, and the lifetime  of the KK resonances  with different $\tilde{b}$ in Tab. \ref{table p4}.
\begin{table*}[htb]
\begin{center}
\begin{tabular}{||c|c|c|c|c|c|c|c||}
\hline
$\tilde{b}$ & $\tilde{\Lambda}_{4}$ &$n$ &parity & $m_{n}^{2}$ &  $P$ & $\Gamma$ & $\tau$    \\
\hline \hline
&-8&1& even & $18.1988$ &$0.372419$ & $0.00810986$ & $123.307$ \\ \cline{2-8}
\raisebox{2.3ex}[0pt]{3}
&-8&2 & odd & $19.8477$ &$0.144147$ & $1.15022$ & $0.8694$ \\ \cline{2-8}
\hline\hline
&-8&1 & odd & $18.0176$ &$0.644363$ & $0.00112908$ & $885.681$ \\ \cline{2-8}
&-8&2 & even & $18.0177$ &$0.336324$ & $0.00112589$ & $888.186$ \\ \cline{2-8}
&-8&3 & odd & $18.0419$ &$0.923572$ & $0.000732362$ & $1365.44$ \\ \cline{2-8}
&-8&4 & even & $18.0424$ &$0.850191$ & $0.000727779$ & $1374.04$ \\ \cline{2-8}
&-8&5 & odd & $18.0763$ &$0.933456$ & $0.000905549$ & $1104.30$ \\ \cline{2-8}
&-8&6 & even & $18.0779$ &$0.94479$ & $ 0.00089571$ & $1116.43$  \\ \cline{2-8}
\raisebox{2.3ex}[0pt]{8}
&-8&7 & odd & $18.1205$ &$0.909459$ & $0.00172875$ & $578.454$ \\ \cline{2-8}
&-8&8 & even & $18.1241$ &$0.89555$ & $0.00141991$ & $587.129$ \\ \cline{2-8}
&-8&9 & odd & $18.1745$ &$0.792877$ & $0.00215697$ & $278.884$ \\ \cline{2-8}
&-8&10  & even & $18.1808$ &$0.755761$ & $0.00358572$ & $404.900$ \\ \cline{2-8}
\hline
\end{tabular}\\
\caption{The influence of the parameter $\tilde{b}$ on the mass spectrum $m_{n}^2$, the relative probability $P$, the width of mass $\Gamma$, and the lifetime $\tau$ of the KK resonances for $\text{AdS}_{4}$ brane ($\tilde{\Lambda}_{4}=-8$).}
\label{table p4}
\end{center}
\end{table*}

\section{Conclusion}\label{Sec4}
In this paper, we investigate the linear tensor perturbation  for $\text{M}_{4}$, $\text{dS}_{4}$ and $\text{AdS}_{4}$ branes  in two-field mimetic gravity. We apply the reconstruction technique to find a set of thick brane solutions in asymptotical $\text{AdS}_{5}$ space-time,  and  derive the master equations for linear tensor perturbations under the TT gauge condition. The Schr\"{o}dinger-like equations for the $\text{M}_{4}$ and $\text{AdS}_{4}$ branes  are factorized into a supersymmetric form; therefore,  the brane systems are stable against the tensor perturbations,  while the $\text{dS}_{4}$ brane  is unstable. For the flat and bent branes, the effective potentials of corresponding Schr\"{o}dinger-like equation behave as  volcano-like or modified-volcano-like  potentials, which may allow a localized zero-mode
responsible for the four-dimensional Newtonian potential and a series of massive resonant
modes. The thick branes have two  parameters  $\tilde{\Lambda}_{4}$ and  $\tilde{b}$, one for the cosmological constant of the bent brane and  the other for the inner structure of the domain wall. We investigate the effect of the two  parameters  on the thick branes, namely,
 \begin{itemize}
\item when $\tilde{\Lambda}_{4}\rightarrow-\tilde{\Lambda}_{4}$, the solution of $\text{dS}_{4}$ changes into that of $\text{AdS}_{4}$. In the limit $\tilde{\Lambda}_{4}\rightarrow0$, the bent brane solution is reduced to the flat brane solution. The number of the resonances does not change with
    $\tilde{\Lambda}_{4}$; however, the masses of resonant KK modes  linearly decrease with the parameter $\tilde{\Lambda}_{4}$ for a fixed $\tilde{b}$.
\item as the parameter $\tilde{b}$ increases, the branes split into sub-branes, and the scalar field $\phi_{1}(\tilde{y})$  changes from a single-kink to a double-kink configuration; the effective potentials  of the extra-dimensional parts of the tensor perturbations also split into multi-wells; the number of gravitational resonance modes increases.
\end{itemize}

Finally, we would like to point out  that the tensor perturbation of the two-field mimetic gravity model is the same as that
of the original single-field mimetic theory \cite{Yi2018} and  GR.   Nevertheless, the two  mimetic scalar fields can generate different  thick branes, leading to new types of effective potential  and graviton resonant modes.

\section*{Acknowledgements}
We sincerely thank Prof. Yu-Xiao Liu for helpful discussions. This work was supported by the National Natural Science Foundation of China (Grants No. 11705070) and the Fundamental Research Fund for Physics of Lanzhou University (No. Lzujbky-2019-ct06). Yi Zhong was supported by the Fundamental Research Funds for the Central Universities (Grants No. 531118010195).

\end{document}